\documentclass[
	aps
	,a4paper
	,pra
	,eqsecnum
	,showpacs
	,floatfix
	,twocolumn
	,merge
	,nofootinbib
	,elide
			]{revtex4-2}
\usepackage{amsmath}
\usepackage{amsfonts}		
\usepackage{graphicx}		
\usepackage{tikz}			
\usepackage{fourier}
\usepackage{amssymb}

\usepackage{calc,tikz}
\definecolor{lime}{HTML}{A6CE39}
\DeclareRobustCommand{\orcidicon}{%
	\begin{tikzpicture}
	\draw[lime, fill=lime] (0,0) 
	circle [radius=0.16] 
	node[white] {{\fontfamily{qag}\selectfont \tiny ID}};
	\end{tikzpicture}
}

\usepackage[%
	unicode=true
	,colorlinks=true
	,linkcolor=blue
	,anchorcolor=blue
	,breaklinks=true
	,citecolor=blue
	,filecolor=cyan
	,menucolor=blue
	,breaklinks=true
	,urlcolor=blue
			]{hyperref}

\begin{document}
\title{
	Magneto-structural dependencies in $3d^2$ systems :
	The trigonal bipyramidal V$^{3+}$ complex
	}

\author{M.~Georgiev\href{https://orcid.org/0000-0003-0598-3360}{\orcidicon}\hspace*{-0.2cm},}
\email{mgeorgiev@issp.bas.bg}
\homepage[\orcidicon]{https://orcid.org/0000-0003-0598-3360}
\affiliation{Institute of Solid State Physics, Bulgarian Academy of Sciences,
	Tsarigradsko Chauss\'ee 72, 1784 Sofia, Bulgaria}

\author{H.~Chamati\href{https://orcid.org/0000-0002-0831-6945}{\orcidicon}\hspace*{-0.2cm},}
\email{chamati@issp.bas.bg}
\homepage[\orcidicon]{https://orcid.org/0000-0002-0831-6945}
\affiliation{Institute of Solid State Physics, Bulgarian Academy of Sciences,
	Tsarigradsko Chauss\'ee 72, 1784 Sofia, Bulgaria}
	
\date{\today}
\begin{abstract} 
We introduce a multi-configurational approach 
to study the magneto-structural correlations in $3d^2$ systems.
The theoretical framework represents a restricted active 
space self-consistent field method, with active space optimized to the 
number of all non-bonding orbitals. 
To demonstrate the validity and effectiveness of the method,
we explore the physical properties of
the trigonal bipyramidal spin-one single-ion magnet 
(C$_6$F$_5$)$_3$trenVCN$^t$Bu.
The obtained theoretical results show a good agreement with the 
experimental data available in the literature. This includes 
measurements for the magnetization, low-field susceptibility, 
cw-EPR and photoluminescence spectroscopy.
The proposed method may be reliably
applied to a variety of $3d^2$ magnetic systems. To this end, and for the sake completeness, we provide detailed analytical and numerical 
representations for the generic Hamiltonian's effective matrix elements related to the crystal 
field, exchange, spin-orbit and Zeeman interactions.
\end{abstract}
\maketitle
	

\section{Introduction}\label{sec:intro}

Being at the frontier between classical and quantum physics, low-dimensional 
magnetic systems 
\cite{bland_ultrathin_2005,cliffe_low-dimensional_2018,zhang_two-dimensional_2021,yang_vi_2020} 
and molecular magnets 
\cite{holynska_2019,sessoli_magnetic_2017,gatteschi_molecular_2006,coronado_molecular_2020} 
continue to generate an ever growing interest among researchers in the
field of magnetism for the last few decades.
Tailoring the underlying quantum features poses a great
challenge for their synthesis, experimental and theoretical
characterization.
Molecular magnets have great scientific and practical potential. 
They have been successfully implemented in resonance imaging 
\protect\cite{zhang_recent_2013,li_molecular_2019,clough_ligand_2019,dutta_magnets_2021,palagi_2021} 
and stand as a promising candidates for building sensor devices \protect\cite{li_spinorbit_2021,liu_mechanochromic_2021,kumar_transition_2020} 
and quantum computing technology
\cite{troiani_molecular_2011,atzori_second_2019,moreno-pineda_2021,hirohata_review_2020} to
name a few.
Despite recent progress \cite{jiang_recent_2021,zhang_recent_2021,wang_recent_2021,yao_recent_2021} 
engineering the magnetic properties of such systems remains a
challenging task and it still requires a lot of 
efforts to pave the route towards their rationalization and 
industrial application.
Let's emphasize that the compound's magnetic behavior 
is tightly related to the type, number and coordination of ligands
with respect to the composing magnetic
centers 
\cite{cao_understanding_2021,gao_molecular_2015,parker_how_2020}.
These has been the subject of extensive interest, here we mention only
some prominent and most recent examples, such as the mononuclear $\beta$-diketonate Dy$^{3+}$ single molecular magnet \cite{xi_regulating_2021},
tetranuclear lanthanide metallocene complexes \cite{mavragani_radicalbridged_2021}, the tetravanadate [V$_{4}$O$_{12}$]$^{4-}$ anion bridged Cu$^{2+}$ complexes \cite{sanchez-lara_magneto-structural_2021}
and the single-ion Co$^{2+}$ one \cite{wu_modulation_2019}.
For an overview on the properties of $3d^2$ type systems the reader may
consult Refs.
\cite{bottcher_ligand_1972,bazhenova_first_2020,jiang_intense_2013,laorenza_tunable_2021}.
Studying the relation between the magnetic properties and the
compound's structure provides a deeper knowledge for the existing
zero-field splitting (ZFS) \cite{rudowicz_concept_1987,*rudowicz_erratum_1988,rudowicz_disentangling_2015}, quantum character of magnetic anisotropy 
and all field dependent properties. It further elucidates the driving mechanisms behind the 
magnetic dynamics that opens up the potential for future applications.

The focus of this study is to reveal the role of the ligand structure
in shaping the
magnetic and spectroscopic properties of $3d^2$ systems and to shed light on the 
relevant contribution of the crystal field, exchange and
spin-orbit interactions. 
To this end, in complement to the quantum perturbation
method, 
we make use of the variational method \cite{cohen_quantum2_2020,QM_Cambridge_2014}.
The associated mathematical framework represents a restrictive active space self-consistent 
field method with an active space spanning all $3d$ orbitals. 
In order to assess the validity of the proposed multi-configurational
approach, we performed a thorough theoretic investigation in the framework of
the 
compound (C$_6$F$_5$)$_3$trenVCN$^t$Bu 
whose magnetic properties have been experimentally probed
\cite{fataftah_trigonal_2020}.
Here, we interpret the experimental data available for 
the magnetization, low-field susceptibility, cw-EPR spectra and photoluminescence.
As the study unfolds, we discuss the role of the crystal field 
and spin-orbit interactions in the magnetization and susceptibility behavior and the 
contribution of the exchange interaction to the absorption and
emission features. 

We would like to emphasize that the proposed approach may be applied to
shed light on the magnetic properties of the whole class of $3d^2$ systems,
regardless of their dimensionality in space -- molecular magnets or
low-dimensional spin systems. Therefore, we provide analytical results
for all interaction terms and effective matrix elements. 

The rest of the paper is structured as follows. 
In Section \ref{sec:theory}, we introduce and elaborate
on the theoretical 
methods used to study the compound under
consideration.
Further, we introduce some basic notations and give explicit representations 
of the applied Hamiltonian and all initial basis sates.
In Sec. \ref{sec:comp}, we provide details of our computational method that
includes representations for all effective matrix elements.
Section \ref{sec:energy} discusses the obtained energy spectrum and its 
dependence on the action of externally applied magnetic field. 
A representation of the obtained ZFS in terms of the
axial ``$D$'' and rhombic ``$E$'' fine structure (FS) parameters is also discussed.
Section \ref{sec:magsub} covers results on the compound's magnetic and spectroscopic properties. 
Finally, a summary of the results is given in Sec.
\ref{sec:Conclusion}. We would like to mention that all numerical results are obtained
with the aid of Wofram Mathematica 12.

\begin{figure}[h!]
	\centering
	\includegraphics[scale=1.1
	]{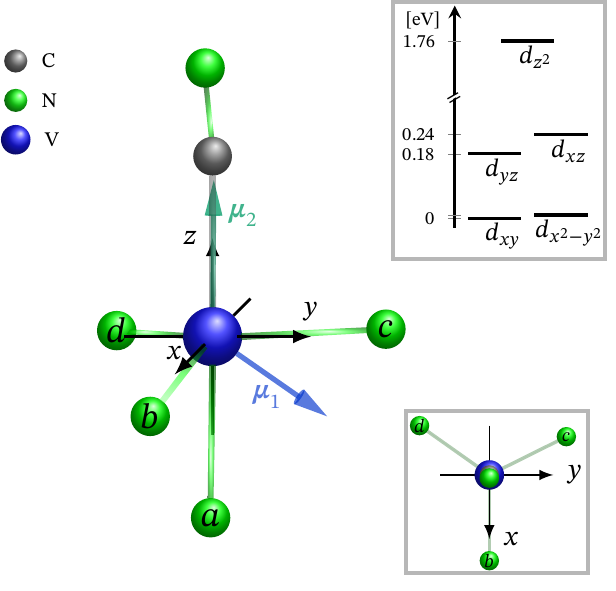}
	\caption{Ball and stick representation of the vanadium ion coordinated by the 
		isocyanide and four nitrogen ligands in the trigonal bipyramidal complex 
		(C$_6$F$_5$)$_3$trenVCN$^t$Bu. The blue, green and gray balls 
		represent the vanadium, nitrogen and carbon elements with coordinates 
		given in Tab. \ref{tab:strpar}. The
		chemical structure of the compound
		with included
		three C$_6$F$_5$ rings, is given in Ref. 
		\cite{fataftah_trigonal_2020}.
		An illustration of the vector associated to the expectation value of the 
		total magnetic moment $\boldsymbol{\mu}_n$,
		corresponding to the ground 
		state $n=1$ and first excited state $n=2$ in the zero-field case is also 
		shown.  
		The inset in the bottom right shows a view from above
		the $xy$ plane,
		while that in top right depicts the respective $d$-orbital energy diagram
		with normalized energy values, such that the lowest one starts from zero. 
	}
	\label{fig:stucture}
\end{figure}

\begin{table}[ht!]
	\caption{The values of the azimuthal angle $\varphi$, polar angle $\vartheta$ and radial 
		distance $\varrho$, for each ligand surrounding the vanadium ion 
		residing at the origin of the reference frame, see Fig. \ref{fig:stucture}.
		All coordinates are calculated with respect to the structural data given in 
		Ref. \cite{fataftah_trigonal_2020}.
		The charge number of each ligand and the metal center 
		relative to both $3d$ electrons is given in the penultimate row. 
		The prefactor related to the spin-orbit interactions
		is given in the last row.
		Their values are fitted to the magnetization, low-field susceptibility, 
		cw-EPR and Photoluminescence experimental 
		data from
		Ref. \cite{fataftah_trigonal_2020}.
	}
	\label{tab:strpar}
	\begin{center}
		\begin{tabular}{lcccccc}
			\hline\hline\noalign{\vspace{1pt}}
			Lig. Number & $1$ & $2$ & $3$ & $4$ & $5$ &  \\
			\noalign{\vspace{1pt}}\hline\noalign{\vspace{1pt}}
			{} & C $\backslash$CN$\backslash$ & N$_a$ & N$_b$ & N$_c$ & N$_d$ & V \\
			\noalign{\vspace{1pt}}\hline\noalign{\vspace{1pt}}
			$\varphi$ [deg] & $0$ & $0$ & $0$ & $116.42$ & $234.74$ & {} \\
			\noalign{\vspace{1pt}}
			$\vartheta$ [deg] & $0$ & $177.02$ & $96.15$ & $97.16$ & $100.49$ & {} \\
			\noalign{\vspace{1pt}}
			$\varrho$ [\AA] & $2.1566$ \ & $2.1447$ \ & $1.9473$ \ & $1.9577$ \ & $1.9619$ & {} \\
			\noalign{\vspace{1pt}}\hline\noalign{\vspace{1pt}}
			$Z_i$ & $\backslash 2.365 \backslash$ \ & $1.157$ \ & $1.157$ \ & $1.157$ \ & $1.157$ \ & $5.9$ \\
			\noalign{\vspace{1pt}}\hline\noalign{\vspace{1pt}}
			$\kappa$ & {} \ & {} \ & {} \ & {} \ & {} \ & 0.43 \\		
			\noalign{\smallskip}\hline\hline
		\end{tabular}
	\end{center}
\end{table}	


\section{Theoretical background}
\label{sec:theory}
\subsection{General considerations}\label{sec:genconsid}

The magnetic and spectroscopic properties of the compound 
(C$_6$F$_5$)$_3$trenVCN$^t$Bu are computed and interpreted using 
a restricted active space self-consistent field approach. The only relativistic 
contribution accounted for in the calculations is the spin-orbit interaction.
In particular, adhering the approximations used in CF theory \cite{maki_ligand_1958,coey_magnetism_2010,miessler_inorganic_2014}, 
we describe both unpaired electrons as localized, to a large extent, around the V$^{3+}$ metal center.
Accordingly, the constructed optimized reaction space is restricted 
to the existing $3d^2$ configuration.
Furthermore, complying to 
the effective field approach used in the Hartree-Fock method \cite{kittel_1987,QM_Cambridge_2014,messiah_quantum_2014},
these electrons are exposed to the average field of all remaining 
``core'' electrons that on average do not contribute neither
to the splitting nor to the 
shifting of energy levels in the ensuing energy spectrum.
Thus, the considered transition metal complex may be
viewed as a good model
of a spin-one system consisting of two $3d$
electrons distributed among six effective point-like
charges. Those are
the vanadium center, the four nitrogen ligands and the isocyanide one.
The five ligands reside on the vertices of a distorted trigonal
bipyramidal structure (see Fig. \ref{fig:stucture}).
We would like to emphasize that in order to overcome the disadvantage
of the localized electron approach 
over the delocalized one used in ligand field theory \cite{miessler_inorganic_2014}, 
we introduce five non-variational parameters, 
one for each ligand, to control
the ligands' electric charge that both $3d$ electrons are exposed to.
Moreover, we have one non-variational parameter to
address the effect of 
electrons' delocalization into the spin-orbit coupling.
The resulting six parameters are then allowed to vary in order to
achieve better agreement between theory and experimental measurements.

We would like to point out, furthermore, that the contributions of 
all orbital and spin magnetic-dipole 
interactions related to  
both $3d$ unpaired electrons and the constituent nuclei are found negligible and
hence omitted from the calculations.
Similarly, the hyperfine interactions 
are also excluded.

Henceforth, we find it convenient to introduce the following
notations: $\mathbf{r}_i=(\alpha_i)_{\alpha\in\mathbb{K}}$ designates the 
$i$-th electron's position vector, where $\mathbb{K}=\{x,y,z\}$ and $i=1,2$.
In spherical coordinates, $\rho_i$, $\theta_i$ and
$\phi_i$ are the $i$-th electron's
radial distance, polar and azimuthal angles, respectively.
By $\hat{\boldsymbol{\mu}}_{l_i}=-\mu_{\mathrm{B}}\hat{\boldsymbol{l}}_i$ 
and 
$\hat{\boldsymbol{\mu}}_{s_i}=-g_e\mu_{\mathrm{B}}\hat{\boldsymbol{s}}_i$,
we denote the corresponding orbital and spin magnetic moment operators, where 
$\mu_{\mathrm{B}}$ is the Bohr magneton, $g_e$ is the electron $g$-factor,
$\hat{\boldsymbol{s}}_i=(\hat{s}^\alpha_i)_{\alpha\in\mathbb{K}}$ and 
$\hat{\boldsymbol{l}}_i=(\hat{l}^\alpha_i)_{\alpha\in\mathbb{K}}$
are the respective spin and angular momentum operators.
The operator of the total magnetic moment $\hat{\boldsymbol{\mu}}$ is 
given by the sum of total spin 
$\hat{\boldsymbol{\mu}}_s=\hat{\boldsymbol{\mu}}_{s_1}+\hat{\boldsymbol{\mu}}_{s_2}$ and 
orbital $\hat{\boldsymbol{\mu}}_l=\hat{\boldsymbol{\mu}}_{l_1}+\hat{\boldsymbol{\mu}}_{l_2}$ ones. 
The position vector of the $i$-th ligand is denoted by 
$\mathbf{d}_i$ with corresponding spherical coordinates 
$\varrho_i$, $\vartheta_i$ and $\varphi_i$, where $i=1,\ldots,5$. 
Here $\vartheta_i$ is the polar and $\varphi_i$ is the azimuthal angles.
Moreover, we set the $z$-axis as the quantization axis and denote the external magnetic field
by $\mathbf{B}=(B_\alpha)_{\alpha\in\mathbb{K}}$ with magnitude $|\mathbf{B}|=B$.
In the bra-ket notation, we use $\bar{1}$ instead of $-1$.
The electric and magnetic constants are denoted as
$\varepsilon$ and $\mu_o$, respectively. The Bohr radius is $r_\mathrm{B}$.
The values of all magnetic moments and their components reported in the text 
are given in units of Bohr magneton.

\subsection{The Hamiltonian}\label{sec:Hamilton}

The Hamiltonian of the considered system reads
\begin{equation}\label{eq:Total_Hamiltonian}
	\hat{H}=\hat{U}_{\mathrm{R}}(\mathbf{r}_1,\mathbf{r}_2) + 
	\hat{U}_{\mathrm{CF}}(\mathbf{r}_1,\mathbf{r}_2) + 
	\hat{U}_{\mathrm{SO}}(\mathbf{r}_1,\mathbf{r}_2) + 
	\hat{U}_{\mathrm{Z}}+\hat{H}_c,
\end{equation}
where $\hat{U}_{\mathrm{R}}$ accounts for the Coulomb repulsion between both 
electrons, $\hat{U}_{\mathrm{CF}}$
represents the interaction of both electrons with the surrounding ligands (crystal field term), 
$\hat{U}_{\mathrm{SO}}$ is the relativistic term that takes into 
account the spin-orbital interactions and $\hat{U}_\mathrm{Z}$
describes the action of externally applied magnetic field. 
The last operator $\hat{H}_c$ on the right-hand-side 
in \eqref{eq:Total_Hamiltonian} is an invariant of the Hamiltonian
system, i.e. $[\hat{H},\hat{H}_c]=0$. 
It produces a constant related to the values of all effective parameters 
that minimize the energy. It is obtained with respect to the variational function 
of all bonding and non-bonding ``core'' electrons. The kinetic terms of both unpaired 
electrons are also accounted
for, since in the case of localized $d$ electrons 
the relevant expectation values are constants.

In spherical coordinates, the series expansion of
the Coulomb term reads \cite{weber_essentials_2014,pisanova_2012}
\begin{equation}\label{eq:Coulomb_term}
	\hat{U}_\mathrm{R}(\mathbf{r}_1,\mathbf{r}_2)\equiv
	\frac{\gamma}{|\mathbf{r}_2-\mathbf{r}_1|}=
\left\lbrace 
\begin{array}{cc}
\displaystyle
	\gamma\sum_{n=0}^{\infty}\frac{\rho^n_2}{\rho^{n+1}_1}P_n(u), & 
	\rho_1>\rho_2, \\ 
[0.5cm]
\displaystyle
	\gamma\sum_{n=0}^{\infty}\frac{\rho^n_1}{\rho^{n+1}_2}P_n(u), & 
	\rho_2>\rho_1,
\end{array}
\right. 
\end{equation}
where $\gamma=e^2/4\pi\varepsilon$ and for all $n$, $P_n(u)$ are the 
Legendre polynomials, with 
$u=\sin\theta_1\sin\theta_2\cos(\phi_1-\phi_2)+\cos\theta_1\cos\theta_2$.

The total CF operator is given by the sum
\begin{equation}\label{eq:CF_term}
	\hat{U}_{\mathrm{CF}}(\mathbf{r}_1,\mathbf{r}_2)=
	\sum_j\big(\hat{U}_{j}(\mathbf{r}_1)+\hat{U}_{j}(\mathbf{r}_2)\big),
\end{equation}
where $j$ runs over the number of all ligands. Here, the series expansion of 
the potential energy accounting for the interaction between the  $k$-th ligand and the $i$-th 
electron is given by \cite{maki_ligand_1958,weber_essentials_2014,coey_magnetism_2010}
\begin{equation}\label{eq:CF_term_1}
	\hat{U}_{k}(\mathbf{r}_i)\equiv
	\frac{\gamma Z_k}{|\mathbf{r}_i-\mathbf{d}_k|}=
	\gamma\sum_{n=0}^{\infty}\frac{\rho^n_i}{\varrho^{n+1}_k}P_n(v_{i,k}), 
\quad
	\varrho_k>\rho_i,
\quad
\forall i, k,
\end{equation}
where, 
$v_{i,k}=\sin\theta_i\sin\vartheta_k\cos(\phi_i-\varphi_k)+\cos\theta_i\cos\vartheta_k$ 
and $Z_k$ is the charge number of the $k$-th ligand to which both $3d$ electrons are exposed.
Notice that $Z_k$, for all $k$, are fitting parameters that allows 
one to distinguish between the different ligands.
Within the used approximations their 
values may not be integers, but their lowest value is unity.

The spin-orbit interaction term is given by
\begin{equation}\label{eq:SO_term}
	\hat{U}_{\mathrm{SO}}(\mathbf{r}_1,\mathbf{r}_2)\equiv
	\frac{g_e\mu_o\mu^2_{\mathrm{B}}}{2\pi}\sum_i\frac{Z}{\rho^3_i}
	\hat{\boldsymbol{l}}_i\cdot\hat{\boldsymbol{s}}_i,
\quad
i=1,2,
\end{equation}
where $Z$ is the charge number of the vanadium metal center with respect 
to both electrons.

Finally the operator describing the
interaction with the externally applied magnetic 
field reads
\begin{equation}\label{eq:Zeeman}
	\hat{U}_{\mathrm{Z}}\equiv
	-\mu_{\mathrm{B}}
	\sum_i\mathbf{B}\cdot\big(\hat{\boldsymbol{l}}_i+g_e\hat{\boldsymbol{s}}_i\big),
\quad
i=1,2.
\end{equation}

\subsection{Initial basis states}

The active space is restricted to the number of all non-bonding orbitals. 
In other words, we have five $3d$ orbitals and two unpaired 
electrons. As a result, we end up with forty five quantum basis states, 
with ten triplets and fifteen singlets. 
Thus, with respect to the exchange symmetry we get 
forty states including only active orbitals,
\begin{subequations}\label{eq:Basis_states}
\begin{equation}\label{eq:Basis_states_a}
\begin{array}{l}
	|\psi_{1,s,m}\rangle\equiv\tfrac{1}{\sqrt{2}}
	\big(|d_{xz},d_{yz}\rangle+(-1)^s|d_{yz},d_{xz}\rangle\big)|s,m\rangle, \\
[0.2cm]
	|\psi_{2,s,m}\rangle\equiv\tfrac{1}{\sqrt{2}}
	\big(|d_{xz},d_{xy}\rangle+(-1)^s|d_{xy},d_{xz}\rangle\big)|s,m\rangle, \\
[0.2cm]
	|\psi_{3,s,m}\rangle\equiv\tfrac{1}{\sqrt{2}}
	\big(|d_{xz},d_{x^2-y^2}\rangle+(-1)^s|d_{x^2-y^2},d_{xz}\rangle\big)|s,m\rangle, \\
[0.2cm]
	|\psi_{4,s,m}\rangle\equiv\tfrac{1}{\sqrt{2}}
	\big(|d_{xz},d_{z^2}\rangle+(-1)^s|d_{z^2},d_{xz}\rangle\big)|s,m\rangle, \\
[0.2cm]
	|\psi_{5,s,m}\rangle\equiv\tfrac{1}{\sqrt{2}}
	\big(|d_{yz},d_{xy}\rangle+(-1)^s|d_{xy},d_{yz}\rangle\big)|s,m\rangle, \\
[0.2cm]
	|\psi_{6,s,m}\rangle\equiv\tfrac{1}{\sqrt{2}}
	\big(|d_{yz},d_{x^2-y^2}\rangle+(-1)^s|d_{x^2-y^2},d_{yz}\rangle\big)|s,m\rangle, \\
[0.2cm]
	|\psi_{7,s,m}\rangle\equiv\tfrac{1}{\sqrt{2}}
	\big(|d_{yz},d_{z^2}\rangle+(-1)^s|d_{z^2},d_{yz}\rangle\big)|s,m\rangle, \\
[0.2cm]
	|\psi_{8,s,m}\rangle\equiv\tfrac{1}{\sqrt{2}}
	\big(|d_{xy},d_{x^2-y^2}\rangle+(-1)^s|d_{x^2-y^2},d_{xy}\rangle\big)|s,m\rangle, \\
[0.2cm]
	|\psi_{9,s,m}\rangle\equiv\tfrac{1}{\sqrt{2}}
	\big(|d_{xy},d_{z^2}\rangle+(-1)^s|d_{z^2},d_{xy}\rangle\big)|s,m\rangle, \\
[0.2cm]
	|\psi_{10,s,m}\rangle\equiv\tfrac{1}{\sqrt{2}}
	\big(|d_{x^2-y^2},d_{z^2}\rangle+(-1)^s|d_{z^2},d_{x^2-y^2}\rangle\big)|s,m\rangle, \\
\end{array}
\end{equation}
and five core orbital states 
\begin{equation}\label{eq:Basis_states_b}
\begin{array}{ll}
	|\psi_{11,0,0}\rangle\equiv|d_{xz},d_{xz}\rangle|0,0\rangle, 
&
	|\psi_{12,0,0}\rangle\equiv|d_{yz},d_{yz}\rangle|0,0\rangle, \\
[0.1cm]
	|\psi_{13,0,0}\rangle\equiv|d_{xy},d_{xy}\rangle|0,0\rangle, 
&
	|\psi_{14,0,0}\rangle\equiv|d_{x^2-y^2},d_{x^2-y^2}\rangle|0,0\rangle, \\
[0.1cm]
	|\psi_{15,0,0}\rangle\equiv|d_{z^2},d_{z^2}\rangle|0,0\rangle,
\end{array}
\end{equation}
\end{subequations}
where $s=0,1$ and $m=\pm s$ are the total spin and spin-magnetic quantum numbers.
Explicit representation of each $d$-state is given in App. \ref{app:d-state}.


\section{Computational details}\label{sec:comp}

\subsection{Coulomb terms}

Calculating the average values of the Coulomb interaction
\eqref{eq:Coulomb_term} up to the $4$-th order in the series expansion
in the basis \eqref{eq:Basis_states}, we obtain zero off-diagonal matrix 
elements.
For the diagonal ones, with
$\langle\psi_{i,s,m}|\hat{U}_{\mathrm{R}}|\psi_{i,s,m}\rangle=E_{i,s,m}$, 
we have
\begin{equation}\label{eq:Matrix_Coulomb}
\begin{array}{ll}
	E_{1,s,m}=V_1+(-1)^s\delta_{1,s}L_1, & E_{2,s,m}=V_4+(-1)^sV_7, \\
	E_{3,s,m}=V_4+(-1)^sV_7, & E_{4,s,m}=V_5+(-1)^sV_8, \\
	E_{5,s,m}=V_4+(-1)^sV_7, & E_{6,s,m}=V_4+(-1)^sV_7, \\
	E_{7,s,m}=V_5+(-1)^sV_8, & E_{8,s,m}=V_2+(-1)^s\delta_{1,s}L_2,  \\
	E_{9,s,m}=V_6+(-1)^sV_9, & E_{10,s,m}=V_6+(-1)^sV_9,  \\
	E_{11,0,0}=V_1+\frac{1}{2}L_1, & E_{12,0,0}=V_1+\frac{1}{2}L_1,  \\
	E_{13,0,0}=V_2+\frac{1}{2}L_2, & E_{14,0,0}=V_2+\frac{1}{2}L_2,  \\
	E_{15,0,0}=V_3, & {} \\	
\end{array}
\end{equation}
where the explicit representations of $\{V_i\}^9_{i=1}$ and 
$\{L_1,L_2\}$, with $Z$ being the only parameter, are provided in 
App. \ref{app:exchange}. The subscript $m$ shows the $3$-fold degeneracy of the
corresponding energy levels.

\subsection{CF terms}

Within the selected active space, the mixing of states by CF is prominent.
In general, we have
$\langle\psi_{i,s,m}|\hat{U}_{\mathrm{CF}}|\psi_{j,s',m'}\rangle
=U_{i,s,m;j,s',m'}$.
For the sake of clarity, we denote the respective diagonal elements by 
$U_{i,s,m}$ to get
\begin{equation}\label{eq:Matrix_CF_diagonal}
\begin{array}{ll}
	U_{1,s,m}=U_{xz}+U_{yz}, & U_{2,s,m}=U_{xz}+U_{xy}, \\ 
	U_{3,s,m}=U_{xz}+U_{x^2-y^2}, & U_{4,s,m}=U_{xz}+U_{z^2},  \\
	U_{5,s,m}=U_{yz}+U_{xy}, & U_{6,s,m}=U_{yz}+U_{x^2-y^2}, \\
	U_{7,s,m}=U_{yz}+U_{z^2}, & U_{8,s,m}=U_{xy}+U_{x^2-y^2}, \\
	U_{9,s,m}=U_{xy}+U_{z^2}, & U_{10,s,m}=U_{x^2-y^2}+U_{z^2}, \\
	U_{11,0,0}=2U_{xz}, & U_{12,0,0}=2U_{yz}, \\
	U_{13,0,0}=2U_{xy}, & U_{14,0,0}=2U_{x^2-y^2}, \\
	U_{15,0,0}=2U_{z^2}. & {}
\end{array}
\end{equation}
The off-diagonal entries are non-zero for all $s'=s$ and $m'=m$. Therefore, 
applying the shorthand notation $U_{i,s,m;j,s,m}\to U_{i,j,s,m}$, we have
\begin{subequations}\label{eq:Matrix_CFoff}
	\begin{equation}\label{eq:Matrix_CFoff_a}
\begin{array}{ll}
	U_{2,1,s,m}=U_{yz,xy}, & U_{3,1,s,m}=U_{yz,x^2-y^2}, \\
	U_{4,1,s,m}=U_{yz,z^2}, & U_{5,1,s,m}=(-1)^sU_{xz,xy},\\
	U_{6,1,s,m}=(-1)^sU_{xz,x^2-y^2}, & U_{7,1,s,m}=(-1)^sU_{xz,z^2}, \\
	U_{3,2,s,m}=U_{xy,x^2-y^2}, & U_{4,2,s,m}=U_{xy,z^2}, \\
	U_{5,2,s,m}=U_{xz,yz}, & U_{8,2,s,m}=(-1)^sU_{xz,x^2-y^2}, \\
	U_{9,2,s,m}=(-1)^sU_{xz,z^2}, & U_{4,3,s,m}=U_{x^2-y^2,z^2}, \\
	U_{6,3,s,m}=U_{xz,yz}, & U_{8,3,s,m}=U_{xz,xy}, \\
	U_{10,3,s,m}=(-1)^sU_{xz,z^2}, & U_{7,4,s,m}=U_{xz,yz}, \\
	U_{9,4,s,m}=U_{xz,xy}, & U_{10,4,s,m}=U_{xz,x^2-y^2}, \\
	U_{6,5,s,m}=U_{xy,x^2-y^2}, & U_{7,5,s,m}=U_{xy,z^2}, \\
	U_{8,5,s,m}=(-1)^sU_{yz,x^2-y^2}, & U_{9,5,s,m}=(-1)^sU_{yz,z^2}, \\
	U_{7,6,s,m}=U_{x^2-y^2,z^2}, & U_{8,6,s,m}=U_{yz,xy}, \\
	U_{10,6,s,m}=(-1)^sU_{yz,z^2}, & U_{9,7,s,m}=U_{yz,xy}, \\
	U_{10,7,s,m}=U_{yz,x^2-y^2}, & U_{9,8,s,m}=U_{x^2-y^2,z^2}, \\
	U_{10,8,s,m}=(-1)^sU_{xy,z^2}, & U_{10,9,s,m}=U_{xy,x^2-y^2}. \\
\end{array}
\end{equation}
Moreover, the average values associated to the singlet only states are given by
\begin{equation}\label{eq:Matrix_CFoff_b}
\begin{array}{ll}
	U_{11,1,0,0}=\sqrt{2} U_{xz,yz}, & U_{12,1,0,0}=\sqrt{2} U_{xz,yz}, \\
	U_{11,2,0,0}=\sqrt{2} U_{xz,xy}, & U_{13,2,0,0}=\sqrt{2} U_{xz,xy}, \\
	U_{11,3,0,0}=\sqrt{2} U_{xz,x^2-y^2}, & U_{14,3,0,0}=\sqrt{2} U_{xz,x^2-y^2}, \\
	U_{11,4,0,0}=\sqrt{2} U_{xz,z^2}, & U_{15,4,0,0}=\sqrt{2} U_{xz,z^2}, \\
	U_{12,5,0,0}=\sqrt{2} U_{yz,xy}, & U_{13,5,0,0}=\sqrt{2} U_{yz,xy}, \\
	U_{12,6,0,0}=\sqrt{2} U_{yz,x^2-y^2}, & U_{14,6,0,0}=\sqrt{2} U_{yz,x^2-y^2}, \\
	U_{12,7,0,0}=\sqrt{2} U_{yz,z^2}, & U_{15,7,0,0}=\sqrt{2} U_{yz,z^2}, \\
	U_{13,8,0,0}=\sqrt{2} U_{xy,x^2-y^2}, & U_{14,8,0,0}=\sqrt{2} U_{xy,x^2-y^2}, \\
	U_{13,9,0,0}=\sqrt{2} U_{xy,z^2}, & U_{15,9,0,0}=\sqrt{2} U_{xy,z^2}, \\
	U_{14,10,0,0}=\sqrt{2} U_{x^2-y^2,z^2}, & U_{15,10,0,0}=\sqrt{2} U_{x^2-y^2,z^2}. \\
\end{array}
\end{equation}
\end{subequations}
The explicit representation of the functions $U_\alpha$ and 
$U_{\alpha,\beta}$, with 
$\alpha\ne\beta\in\{xz,yz,xy,x^2-y^2,z^2\}$, is shown in App. \ref{app:CF}.
We would like to point out that all functions in \eqref{eq:Matrix_CFoff} are 
real, with  parameters being the ligands' coordinates,
$Z_1,\ldots,Z_5$ and $Z$.
The computations are performed up to the forth order in \eqref{eq:CF_term_1}.

\subsection{Spin-orbit terms}\label{sec:SO}

All diagonal matrix elements associated 
to the total spin-orbit operator
\eqref{eq:SO_term} are zero. For both electrons, the corresponding
spin-orbit coupling 
$\zeta=\langle g_e\mu_o\mu^2_{\mathrm{B}}Z/2\pi\rho^3_i \rangle=
(35.604\times10^{-4})Z^4$ meV.
In order to write down the respective matrix elements in a
transparent way, we 
introduce the parameter $\eta=\kappa\zeta/2\sqrt{2}$, with 
$0\le\kappa\le1$. 
The prefactor $\kappa$ is included to accounts for a possible delocalization of 
both electrons. In other words, if $\kappa\approx1$, then the initial assumption 
of localized to a large extent electrons holds. On the other hand, 
if $\kappa<1$, it would be a 
sign that the considered electrons may be partially delocalized.
Now, for $\langle\psi_{i,s,m}|\hat{U}_{\mathrm{SO}}|\psi_{j,s',m'}\rangle
=Y_{i,s,m;j,s',m'}$, we get
\begin{equation*}\label{eq:Matrix_SO}
\begin{array}{ll}
	Y_{2,1,\pm1;1,s,0}=(\pm1)^s\eta, & 
	Y_{3,1,\pm1;1,s,0}=\pm i(\pm1)^{s}\eta, \\ 
	Y_{4,1,\pm1;1,s,0}=\pm i\sqrt{3}(\pm1)^{s}\eta, & 
	Y_{5,1,\pm1;1,s,0}=-i(\mp1)^s\eta, \\
	Y_{6,1,\pm1;1,s,0}=(\mp1)^s\eta, &	
	Y_{7,1,\pm1;1,s,0}=-\sqrt{3}(\mp1)^s\eta, \\
	Y_{1,1,\pm1;2,s,0}=-(\pm1)^s\eta, & 
	Y_{3,1,\pm1;2,1,\pm1}=\mp i2\sqrt{2}\eta, \\
	Y_{5,1,\pm1;2,1,\pm1}=\pm i\sqrt{2}\eta, & 
	Y_{8,1,\pm1;2,s,0}=(\mp)^s\eta, \\ 
	Y_{9,1,\pm1;2,s,0}=-\sqrt{3}(\mp1)^s\eta, &
	Y_{1,1,\pm1;3,s,0}=\mp(\pm1)^{s}\eta, \\
	Y_{6,1,\pm1;3,1,\pm1}=\pm i\sqrt{2}\eta, & 
	Y_{8,1,\pm1;3,s,0}=\pm i(\mp1)^{s}\eta, \\
	Y_{10,1,\pm1;3,s,0}=-\sqrt{3}(\mp1)^{s}\eta, & 
	Y_{1,1,\pm1;4,s,0}=\mp i\sqrt{3}(\pm1)^s\eta, \\
	Y_{7,1,\pm1;4,1,\pm1}=\pm i\sqrt{2}\eta, & 
	Y_{9,1,\pm1;4,s,0}= \pm i(\mp1)^{s}\eta, \\	
	Y_{10,1,\pm1;4,1,0}=-(\mp1)^s\eta, & 
	Y_{1,1,\pm1;5,s,0}=\mp i(\pm1)^{s}\eta, \\
	Y_{6,1,\pm1;5,1,\pm1}=\mp i2\sqrt{2}\eta, &
	Y_{8,1,\pm1;5,s,0}=\pm i(\mp1)^{s}\eta, \\
	Y_{9,1,\pm1;5,s,0}=\pm i\sqrt{3}(\mp1)^s\eta, &
	Y_{1,1,\pm1;6,s,0}= (\pm1)^s\eta, \\
	Y_{8,1,\pm1;6,s,0}= -(\mp1)^s\eta, & 
	Y_{10,1,\pm1;6,s,0}= \pm i\sqrt{3}(\mp 1)^s\eta \\
\end{array}
\end{equation*} 
and
\begin{equation*}
\begin{array}{ll}
	\displaystyle 
	Y_{1,1,\pm1;7,s,0}= -\sqrt{3}(\pm1)^s\eta, & 
	Y_{9,1,\pm1;7,s,0}= -(\mp1)^s\eta, \\
	Y_{10,1,\pm1;7,s,0}=\mp i(\mp1)^{s}\eta, & 
	Y_{2,1,\pm1;8,s,0}= (\pm1)^s\eta, \\
	Y_{3,1,\pm1;8,s,0}=\mp i(\mp1)^{s}\eta, & 
	Y_{5,1,\pm1;8,s,0}=\pm i(\pm1)^{s}\eta, \\
	Y_{6,1,\pm1;8,s,0}= (\mp1)^{s}\eta, & 
	Y_{2,1,\pm1;9,s,0}= -\sqrt{3}(\pm1)^{s}\eta, \\
	Y_{4,1,\pm1;9,s,0}=\mp i(\mp1)^{s}\eta, & 
	Y_{5,1,\pm1;9,s,0}=\pm i\sqrt{3}(\pm1)^{s}\eta, \\
	Y_{7,1,\pm1;9,s,0}= (\mp1)^{s}\eta, & 
	Y_{10,1,\pm1;9,1,\pm1}=\mp i2\sqrt{2}\eta, \\
	Y_{3,1,\pm1;10,s,0}=-\sqrt{3}(\pm1)^{s}\eta, & 
	Y_{4,1,\pm1;10,s,0}=(\mp1)^{s}\eta, \\
	Y_{6,1,\pm1;10,s,0}=\pm i\sqrt{3}(\pm1)^{s}\eta, & 
	Y_{7,1,\pm1;10,s,0}=\pm i(\pm1)^{s}\eta. \\
\end{array}
\end{equation*}
Further, the spin-orbit interactions mix the single orbital states 
\eqref{eq:Basis_states_b} with the remaining ones, such that
\begin{equation*}
\begin{array}{lll}
	Y_{2,1,\pm1;11,0,0}=\mp i\sqrt{2}\eta, & 
	Y_{3,1,\pm1;11,0,0}=\sqrt{2}\eta, \\
	Y_{4,1,\pm1;11,0,0}=-\sqrt{6}\eta, & 
	Y_{5,1,\pm1;12,0,0}=\sqrt{2}\eta, \\
	Y_{6,1,\pm1;12,0,0}=\pm i\sqrt{2}\eta, & 
	Y_{7,1,\pm1;12,0,0}=\pm i\sqrt{6}\eta, \\
	Y_{2,1,\pm1;13,0,0}=\mp i\sqrt{2}\eta, & 
	Y_{5,1,\pm1;13,0,0}=\sqrt{2}\eta, \\
	Y_{3,1,\pm1;14,0,0}=\sqrt{2}\eta, & 
	Y_{6,1,\pm1;14,0,0}=\pm i\sqrt{2}\eta, \\
	Y_{4,1,\pm1;15,0,0}=-\sqrt{6}\eta, & 
	Y_{7,1,\pm1;15,0,0}=\pm i\sqrt{6}\eta. \\
\end{array}
\end{equation*}

\subsection{Zeeman terms}\label{sec:zeeman}

Due to the symmetry of the orbitals in \eqref{eq:Basis_states}, the  
diagonal matrix elements obtained from \eqref{eq:Zeeman}, 
$Z_{i,s,m}=\langle\psi_{i,s,m}|\hat{U}_{\mathrm{Z}}|\psi_{i,s,m}\rangle$, 
depend only on the total spin. 
Thus, we have $Z_{i,s,m}=g_e\mu_{\mathrm{B}}mB_z$. 
Furthermore, since we work in the basis of the spin $z$ component, there are forty 
spin-only off-diagonal elements corresponding to 
the mixing between $m=0$ and $m=\pm1$ states. For all $i=1,\ldots,10$, the 
non-conjugate ones read $Z_{i,1,\pm1;i,1,0}=-g_e\mu_{\mathrm{B}}(B_x\mp iB_y)/\sqrt{2}$.
The remaining off-diagonal elements depend exclusively on the mixing of orbitals.
Using the substitution $Z_{i,s,m;j,s,m}\to Z_{i,j,s,m}$, 
with respect to the states \eqref{eq:Basis_states_a}, we obtain
\begin{equation*}
\begin{array}{ll}
	Z_{2,1,s,m}=-i\mu_{\mathrm{B}}B_y, & 
	Z_{3,1,s,m}=-i\mu_{\mathrm{B}}B_x, \\
	Z_{4,1,s,m}=-i\sqrt{3}\mu_{\mathrm{B}}B_x, & 
	Z_{5,1,s,m}=i(-1)^{s}\mu_{\mathrm{B}}B_x, \\
	Z_{6,1,s,m}=-i(-1)^{s}\mu_{\mathrm{B}}B_y, & 
	Z_{7,1,s,m}=i(-1)^{s}\sqrt{3}\mu_{\mathrm{B}}B_y, \\
	Z_{3,2,s,m}=i2\mu_{\mathrm{B}}B_z, & 
	Z_{5,2,s,m}=-i\mu_{\mathrm{B}}B_z, \\
	Z_{8,2,s,m}=-i(-1)^s\mu_{\mathrm{B}}B_y, & 
	Z_{9,2,s,m}=i(-1)^s\sqrt{3}\mu_{\mathrm{B}}B_y, \\
	Z_{6,3,s,m}=-i\mu_{\mathrm{B}}B_z, & 
	Z_{8,3,s,m}=i\mu_{\mathrm{B}}B_x, \\
	Z_{10,3,s,m}=i(-1)^s\sqrt{3}\mu_{\mathrm{B}}B_y, & 
	Z_{7,4,s,m}=-i\mu_{\mathrm{B}}B_z, \\
	Z_{9,4,s,m}=i\mu_{\mathrm{B}}B_x, & 
	Z_{10,4,s,m}=-i\mu_{\mathrm{B}}B_y, \\
	Z_{6,5,s,m}=i2\mu_{\mathrm{B}}B_z, & 
	Z_{8,5,s,m}=-i(-1)^s\mu_{\mathrm{B}}B_x, \\
	Z_{9,5,s,m}=-i(-1)^s\sqrt{3}\mu_{\mathrm{B}}B_x, & 
	Z_{8,6,s,m}=-i\mu_{\mathrm{B}}B_y, \\
	Z_{10,6,s,m}=-i(-1)^s\sqrt{3}\mu_{\mathrm{B}}B_x, & 
	Z_{9,7,s,m}=-i\mu_{\mathrm{B}}B_y, \\
	Z_{10,7,s,m}=-i\mu_{\mathrm{B}}B_x, & 
	Z_{10,9,s,m}=i2\mu_{\mathrm{B}}B_z \\
\end{array}
\end{equation*}
and with the consideration of singlet states \eqref{eq:Basis_states_b}, we get
\begin{equation*}
\begin{array}{ll}
	Z_{11,1,0,0}=i\sqrt{2}\mu_{\mathrm{B}}B_z, & 
	Z_{12,1,0,0}=-i\sqrt{2}\mu_{\mathrm{B}}B_z, \\
	Z_{11,2,0,0}=-i\sqrt{2}\mu_{\mathrm{B}}B_x, & 
	Z_{13,2,0,0}=i\sqrt{2}\mu_{\mathrm{B}}B_x, \\
	Z_{11,3,0,0}=i\sqrt{2}\mu_{\mathrm{B}}B_y, & 
	Z_{14,3,0,0}=-i\sqrt{2}\mu_{\mathrm{B}}B_y, \\	
	Z_{11,4,0,0}=-i\sqrt{6}\mu_{\mathrm{B}}B_y, & 
	Z_{15,4,0,0}=i\sqrt{6}\mu_{\mathrm{B}}B_y, \\
	Z_{12,5,0,0}=i\sqrt{2}\mu_{\mathrm{B}}B_y, & 
	Z_{13,5,0,0}=-i\sqrt{2}\mu_{\mathrm{B}}B_y, \\
	Z_{12,6,0,0}=i\sqrt{2}\mu_{\mathrm{B}}B_x, & 
	Z_{14,6,0,0}=-i\sqrt{2}\mu_{\mathrm{B}}B_x, \\
	Z_{12,7,0,0}=i\sqrt{6}\mu_{\mathrm{B}}B_x, & 
	Z_{15,7,0,0}=-i\sqrt{6}\mu_{\mathrm{B}}B_x, \\
	Z_{13,8,0,0}=-i2\sqrt{2}\mu_{\mathrm{B}}B_z, & 
	Z_{14,8,0,0}=i2\sqrt{2}\mu_{\mathrm{B}}B_z. \\	
\end{array}
\end{equation*}
%


\section{Energy spectrum}\label{sec:energy}

\subsection{Zero-field spectrum}\label{sec:zfs}

The energy spectrum of the considered vanadium complex is obtained
after direct 
diagonalization of the total matrix given by the sum of 
the four matrices corresponding to the interactions in
Hamiltonian
\eqref{eq:Total_Hamiltonian} and with elements given in Sec. \ref{sec:comp}.
The spectrum is non-degenerate and hence it is build up of forty five energy levels for 
both cases $B=0$ and $B\ne0$. 
Since the value of the ground state energy does not affect the final 
splitting and shifting of levels, the respective eigenvalues are normalized such 
that the ground state energy equals zero. 
The resulting energy level sequence depends on two sets of parameters. On one 
hand we have the ligands' coordinates given in Tab. \ref{tab:strpar} and on the 
other the fitting parameters $\kappa$, $Z$ and $Z_k$, for $k=1,\ldots,5$.
The relevant structural parameters are experimentally determined and provided
in Ref. \cite{fataftah_trigonal_2020}.
The values of all ``$Z$'' parameters are given in the penultimate row 
of Tab. \ref{tab:strpar}
and in conjunction with $\kappa=0.43$ their values are fitted in accordance to the  
low-field susceptibility, magnetization, cw-EPR experimental 
and Photoluminescence data from Ref. \cite{fataftah_trigonal_2020}.

\begin{figure}[h!]
	\centering
	\includegraphics[scale=0.95]{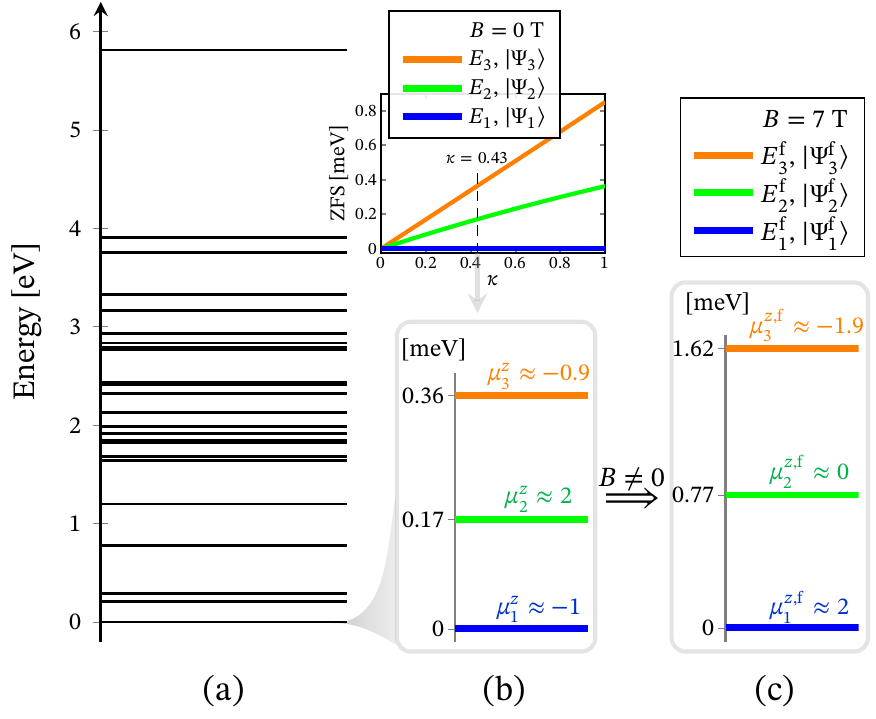}
	\caption{Energy spectrum of the complex (C$_6$F$_5$)$_3$trenVCN$^t$Bu.
		The energy levels are shifted such 
		that the ground state one is zero. 
		(a) Represents all energy spectra consisting of forty five levels. 
		(b) Depicts the first three energy levels governing the compound's 
		magnetic properties in the zero-field case with parameters given 
		in Tab. \ref{tab:strpar}.
		It further shows the splitting of these
		levels as a function $\kappa$.
		(c) Shows the shifting of the first three levels under the action of 
		an externally applied magnetic field along the $z$ axis of the 
		compounds reference frame, see Fig. \ref{fig:stucture}, 
		with magnitude $B\equiv B_z=7$ T. The blue, 
		green and orange lines indicate the ground, first and second excited 
		energy levels, respectively.
		The corresponding expectation values of the
		$z$-component of the total magnetic moment
		in [$\mu_{\mathrm{B}}$] are also given.
		The respective designations in the case of absence
		and presence of 
		magnetic field are $\mu^z_n$ and 
		$\mu^{z,\mathrm{f}}_n$, $n=1,2,3$. 
		The superscript ``f'' marks the selected magnetic 
		field direction and magnitude.
	}
	\label{fig:en}
\end{figure}

The energy spectrum in the absence of an external magnetic field is shown on Fig. 
\ref{fig:en} (a). The value of each energy level is provided in App. 
\ref{app:enspectrum}. Fig. \ref{fig:en} (b) depicts the first 
three energy levels, which represent the obtained ZFS \cite{abragam_electron_2012,bencini_electron_1990,pilbrow_transition_1990,misra_multifrequency_2011,rudowicz_disentangling_2015}
and its dependence on $\kappa$. 

The combined effect of spin-orbital interactions and distorted trigonal 
bipyramidal geometry, see Fig. \ref{fig:stucture}, determines
the ground state as a mixing involving only $m=0$ and $m=-1$ spin
states, i.e.
\begin{align}\label{eq:ZGstate}
	|\Psi_1\rangle=&
	(-0.14+0.14i) |\psi_{1,1,0}\rangle-(0.14+0.15i)|\psi_{1,1,\bar{1}}\rangle 
\nonumber \\ & 
	+(0.25-0.24i)|\psi_{2,1,0}\rangle + (0.25+0.27i) |\psi_{2,1,\bar{1}}\rangle 
\nonumber \\ & 
	-(0.13-0.13i)|\psi_{3,1,0}\rangle - (0.13+0.14i) |\psi_{3,1,\bar{1}}\rangle 
\nonumber \\ & 
	-(0.24-0.23i)|\psi_{5,1,0}\rangle - (0.25+0.25i) |\psi_{5,1,\bar{1}}\rangle 
\nonumber \\ & 
	+(0.29-0.28i)|\psi_{6,1,0}\rangle + (0.29+0.31i)|\psi_{6,1,\bar{1}}\rangle 
\nonumber \\ & 
- (0.065 - 0.063i)|\psi_{8,1,0}\rangle - (0.066 + 0.071i)|\psi_{8,1,\bar{1}}\rangle
\nonumber \\ &
+ \sum_{n\ge2} O\big(10^{-n}\big)|\psi_{\ldots}\rangle.
\end{align}
Therefore, we have a probability of approximately $47\%$ to observe the system 
in $m=0$ state and $53\%$ in a state with $m=-1$. Consequently, with 
expectation values given in Bohr magneton units, for the total magnetic moment we have 
$\boldsymbol{\mu}_1=(-0.0171, 1.4091, -1.0865)$. The orbital contribution 
is negligible, thus
$\boldsymbol{\mu}_{s,1}=(-0.0181, 1.4103, -1.0612)$
and $\boldsymbol{\mu}_{l,1}=(0.001, -0.0011, -0.0253)$.

The distortion in coordination geometry also determines the first excited state 
as a superposition of only $m=1$ spin states
\begin{align}\label{eq:ZExstate}
	|\Psi_2\rangle=&
	(0.19-0.21i)|\psi_{1,1,1}\rangle - (0.34-0.37i)|\psi_{2,1,1}\rangle 
\nonumber \\ & 	
	+ (0.17-0.20i)|\psi_{3,1,1}\rangle + (0.33-0.35i)|\psi_{5,1,1}\rangle 
\nonumber \\ & 
	- (0.39-0.44i)|\psi_{6,1,1}\rangle + (0.089-0.098i)|\psi_{8,1,1}\rangle
\nonumber \\ & 
	+ \sum_{n\ge2} O\big(10^{-n}\big)|\psi_{\ldots}\rangle,
\end{align}
where $\boldsymbol{\mu}_2=(0.0157, 0.0037, 2.0509)$ and  
$\boldsymbol{\mu}_{s,2}=(0, 0, 1.9994)$.
Thus, any excitation to $E_2$ would be 
associated to an almost fully polarized magnetic moment.
An illustration of both vectors $\boldsymbol{\mu}_1$ and $\boldsymbol{\mu}_2$ 
is shown on Fig. \ref{fig:stucture}.

The second excited state, $|\Psi_3\rangle$, is given by the same superposition 
of initial basis states as the ground state \eqref{eq:ZGstate}, but with 
different probability coefficients values, 
see \eqref{eq:Ap_ZExstate_2}. The corresponding total magnetic moment is 
$\boldsymbol{\mu}_3=(0.0036, -1.4241, -0.9634)$, with contribution 
from the spin component
$\boldsymbol{\mu}_{s,3}=(0.0182, -1.4105, -0.9379)$.

\begin{figure}[ht!]
	\centering
	\includegraphics[scale=1]{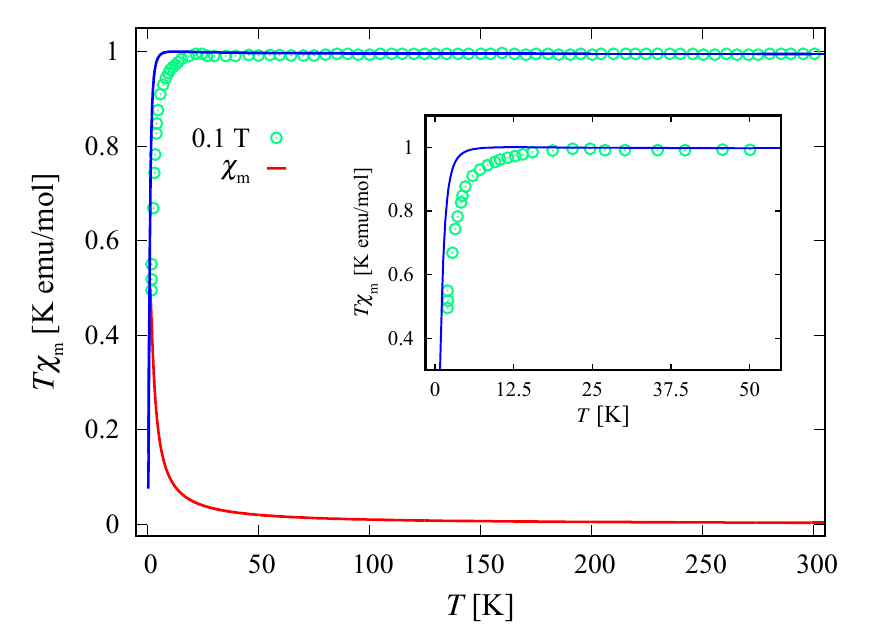}
	\caption{Molar susceptibility of (C$_6$F$_5$)$_3$trenVCN$^t$Bu multiplied
		by the temperature as a function of the temperature. The experimental 
		data, taken from Ref. \cite{fataftah_trigonal_2020},
		are depicted by green 
		circles. The computational results are represented by
		a solid blue line.
		For the sake of completeness the calculated low-field molar susceptibility is shown 
		by a solid red line. All relevant parameters are given in Tab. 
		\ref{tab:strpar}.
	}
	\label{fig:TST}
\end{figure}

\begin{figure}[ht!]
	\centering
	\includegraphics[scale=1]{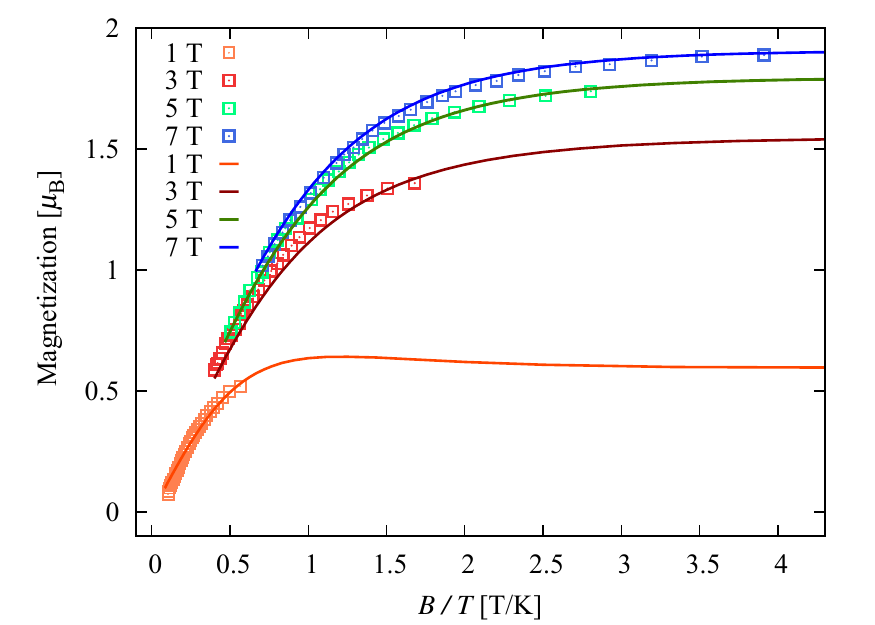}
	\caption{Magnetization plot of the compound (C$_6$F$_5$)$_3$trenVCN$^t$Bu 
		at $1$, $3$, $5$ and $7$ T magnetic field values.
		The experimental data provided by Ref.
		\cite{fataftah_trigonal_2020} are
		depicted by empty squares. The solid curves represent the theoretical 
		results discussed in Sec. \ref{sec:magsub}. The values of all corresponding 
		parameters are given in Tabs. \ref{tab:strpar} and \ref{tab:Magfield}.
	}
	\label{fig:Mag}
\end{figure}

It is essential to emphasize that the first almost total singlet state, with 
approximately $99 \%$ probability to observe $s=0$, is
\begin{align}\label{eq:S26}
	|\Psi_{26}\rangle=&
 0.092|\psi_{11,0,0}\rangle-0.24|\psi_{12,0,0}\rangle-0.34|\psi_{13,0,0}\rangle
\nonumber \\ & 
+0.18|\psi_{14,0,0}\rangle - 0.26|\psi_{2,0,0}\rangle + 0.31|\psi_{3,0,0}\rangle 
\nonumber  \\ &	 
+ 0.74|\psi_{5,0,0}\rangle - 0.11|\psi_{6,0,0}\rangle - 0.23|\psi_{8,0,0}\rangle
\nonumber  \\ &
+ \sum_{n\ge2} O\big(10^{-n}\big)|\psi_{\ldots}\rangle.
\end{align}
The corresponding energy level lies very high in the zero-temperature energy 
spectrum shown on Fig. \ref{fig:en} (a). It is approximately $1.846$ eV. Such a 
large exchange coupling is typical for transition metal complexes with 
localized around the metal 
center electrons and suggests a possible phosphorescence observation.
On the other hand, the small energy gaps $\Delta E_{21}\approx0.17$ meV and 
$\Delta E_{31}\approx0.36$ meV, make the zero-field and low-field 
magnetic properties of the considered complex highly sensitive to the
variations in temperature. This feature is clearly demonstrated by the 
temperature dependence of the susceptibility depicted on Fig. \ref{fig:TST} 
and the magnetization behavior shown on  
Fig. \ref{fig:Mag} at $B=1$ T. This point will be discussed below in Sec. \ref{sec:magsub}.

\begin{table}[!ht]
	\caption{Magnetization ($4$-th row) for given magnetic field values $B$ and 
		temperatures to the saturation values ($1$-st row).
	The last row indicates the variation of the effective angle
	``$\nu$" between the vectors of the applied magnetic 
	field and the magnetization with the change of $B$.
	For comparison, the third row shows the $z$ component of the molecule's
	magnetic moment at the same temperature and magnetic field values,
	where in this case $\nu$ represents the angle between the magnetic field and $z$ 
	axis of the compound's reference frame.  
		\label{tab:Magfield} }
	\begin{center}
		\begin{tabular}{lccccc}
			\hline\hline\noalign{\vspace{1pt}}
			$T$ [K] & $\le0.1$ & $0.2$ & $0.6$ & $1$ & $1.35$  \\ 
			\noalign{\vspace{1pt}}\hline\noalign{\vspace{1pt}}
			$B$ [T] & $0$ & $1$ & $3$ & $5$ & $7$  \\ 
			\noalign{\vspace{1pt}}\hline\noalign{\vspace{1pt}}
			$m_z$ [$\mu_{\mathrm{B}}$] & $-1.0865$ &  $-0.4666$ & $1.3382$ &  $1.7712$ & $1.9240$   \\
			\noalign{\vspace{1pt}}
			$M$ [$\mu_{\mathrm{B}}$] & $0$ &  $0.5952$ & $1.5434$ & $1.7920$ & $1.9046$  \\
			\noalign{\vspace{1pt}}\hline\noalign{\vspace{1pt}}
			$\nu$ [deg] & \phantom{sss} &  $47.4681$ & $34.8129$ &  $24.6142$ & $16.9022$   \\
			\noalign{\smallskip}\hline\hline
		\end{tabular}
	\end{center}
\end{table}	

According to all approximations taken into account in Sec. \ref{sec:genconsid}, 
the obtained ZFS results due to the spin-orbit interactions acting within the CF basis.
Therefore, any consideration with $\kappa\to0$ leads 
a 3-fold degenerate ground state. For further details see Fig. \ref{fig:en} (b).
On the other hand, for $\kappa\to1$ the ZFS increases to the extent that the theory is 
no longer able to reproduce the overall experimental observations.
To gain insight into the interrelationship between the two physically different notions: ZFS and CF, 
we recommend the interested reader to consult Refs. \cite{rudowicz_concept_1987,*rudowicz_erratum_1988,rudowicz_disentangling_2015}.

\subsection{The magnetic field effect}\label{sec:MagFieldEffect}

The interaction of the considered complex with the externally applied magnetic field
is taken into account by the Zeeman effective matrix with elements given in 
Sec. \ref{sec:zeeman}. The 
zero-field spectrum is not degenerate with respect to $m$, due to the
distorted geometry, and so no Zeeman splitting can be
observed. As a consequence, we witness only shifting of the energy
levels.
However, the change of the ground state takes place in a way that resembles the 
Zeeman effect.

An example of how the energy levels shift under the action of the externally applied 
magnetic field, for $B_z\equiv B=7$ T, is demonstrated on Fig. \ref{fig:en} (c). 
The superscript ``f'' indicates the selected magnetic 
field direction and the magnitude.
The resulting shifting is related to a change of the eigenstates.
The second excited state from the zero-field spectrum
\eqref{eq:ZExstate} is now the ground state.
Thus, we have $|\Psi_2\rangle\to|\Psi^\mathrm{f}_1\rangle$, where the variation in the 
probability coefficients value is negligible, such that 
$\boldsymbol{\mu}_2\to\boldsymbol{\mu}^{\mathrm{f}}_1=(0.015, 0.0048, 2.0288)$.
This is not the case with the first excited state $|\Psi^\mathrm{f}_2\rangle$, 
which is a transformation of the zero-field ground state
\eqref{eq:ZGstate}. We get
\begin{align}\label{eq:FEXstate}
	|\Psi^\mathrm{f}_2\rangle=&
	(0.18 - 0.21 i) |\psi_{1,1,0}\rangle+(0.048 + 0.043i)|\psi_{1,1,\bar{1}}\rangle 
	\nonumber \\ & 
	-(0.32 - 0.37i)|\psi_{2,1,0}\rangle - (0.081 + 0.074i) |\psi_{2,1,\bar{1}}\rangle 
	\nonumber \\ & 
	+(0.17 - 0.19i)|\psi_{3,1,0}\rangle + (0.040 + 0.039i) |\psi_{3,1,\bar{1}}\rangle 
	\nonumber \\ & 
	+(0.31 - 0.36 i)|\psi_{5,1,0}\rangle + (0.081 + 0.069i) |\psi_{5,1,\bar{1}}\rangle 
	\nonumber \\ & 
	-(0.38 - 0.43i)|\psi_{6,1,0}\rangle - (0.096 + 0.086i)|\psi_{6,1,\bar{1}}\rangle 
	\nonumber \\ & 
	+ (0.084 - 0.098i)|\psi_{8,1,0}\rangle + (0.022 +0.020i)|\psi_{8,1,\bar{1}}\rangle
	\nonumber \\ &
	+ \sum_{n\ge2} O\big(10^{-n}\big)|\psi_{\ldots}\rangle.
\end{align}
Hence, the orbital and spin terms given in Sec. \ref{sec:zeeman}
enhance the 
probability of observing the $m=0$ state from $47\%$ at $B=0$ T to 
approximately $95\%$ in the considered case. 
At the same time the value of corresponding energy level increases and 
$E_1\to E^{\mathrm{f}}_2$.
Respectively, the expectation values of the total magnetic moment also change. 
We get $\boldsymbol{\mu}_1\to\boldsymbol{\mu}^{\mathrm{f}}_2=(0.0022, 0.6102, -0.1193)$. 

The second excited state $|\Psi^\mathrm{f}_3\rangle$ is a transformation of 
$|\Psi_3\rangle$, and it favors the $m=-1$ outcome with approximately 
$94.8\%$ probability rather than $46.4\%$ obtained for the zero-field case. 
As a result, we get 
$\boldsymbol{\mu}_3\to\boldsymbol{\mu}^{\mathrm{f}}_3=(0.0011, -0.6053, -1.9755)$.
For the explicit representation of the second excited quantum state,
see e.g. \eqref{eq:Ap_Exstate_2}.

We would like to point out that the observed shifting of energy levels and the 
absence of a genuine Zeeman splitting in the present study is similar
to the same feature found for the magnetic behavior of 
Ni$_{4}$Mo$_{12}$ molecular magnet 
\cite{georgiev_magnetization_2020} and Er$^{3+}$ complexes 
\cite{georgiev_erbium_2021},
except that for the latter compounds such a dependence of the fine
structure on an externally 
applied magnetic field is more complex and it is tightly related to the
exchange interactions \cite{georgiev_molecular_2020}.

\subsection{Orbital-energy diagram}

In addition to the energy spectrum we introduce the splitting of the $3d$ orbitals.
The corresponding diagram is depicted on the top right in Fig. \ref{fig:stucture}.
Each single electron $3d_\alpha$ orbital has an energy $U_\alpha$ given in App. \ref{app:CF}, where 
$\alpha\in\{xz,yz,xy,x^2-y^2,z^2\}$.
As expected, due to the presence of CN ligand the $d_{z^2}$ orbital 
is the highest in energy.
For the same reason $d_{xz}$ and $d_{yz}$ orbitals are higher in energy
than the $d_{xy}$ and $d_{x^2-y^2}$ ones, which is opposite to the case of 
identical ligands.

\begin{figure}[h!]
	\centering
	\includegraphics[scale=1]{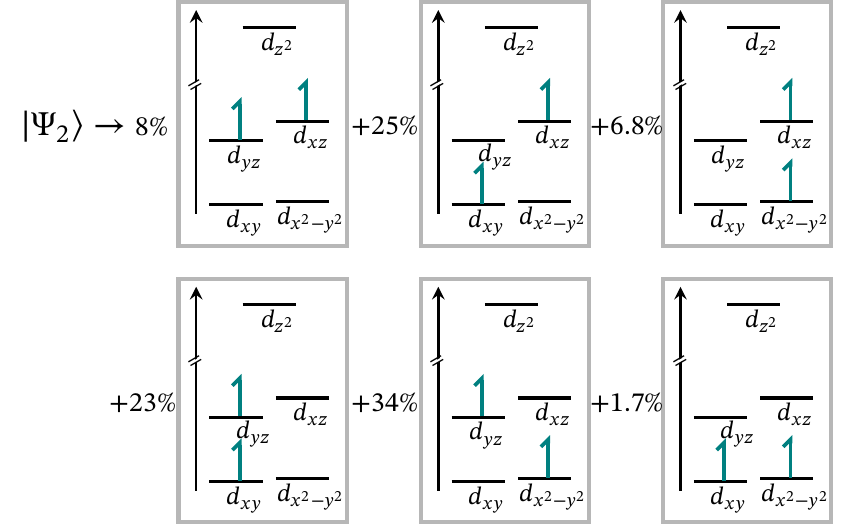}
	\caption{Sketch representing the first excited state  
		in the zero-field energy spectrum as a superposition of spin-orbital 
		diagrams. The corresponding energy scale is depicted on Fig. \ref{fig:stucture}.
		The diagrams' arrangement is given in accordance to the superposition 
		of the configuration state functions \eqref{eq:ZExstate}, 
		with only the first six states included.
		The coefficients are replaced by the corresponding probabilities, 
		given in percentages.
		The individual spins with their mutual orientations are illustrated as 
		green half-arrows.
	}
	\label{fig:Superpos}
\end{figure}

Each eigenstate can be schematically represented as a superposition of 
such diagrams with included spin-configurations.
For example, the first excited state given in \eqref{eq:ZExstate} is
the superposition of six spin-orbital configurations shown on Fig. 
\ref{fig:Superpos}.
The approximate probability of observing each one is given in percentage.

\subsection{The determination of the ZFS parameters $D$ and $E$}\label{sec:DE}

Since we have neglected all dipole-dipole magnetic interactions the obtained FS 
shown on Figs. \ref{fig:en} (b) is entirely determined from the spin-orbit interactions 
within CF basis states \eqref{eq:Basis_states}.
Thus, quantifying the ZFS in terms of the corresponding parameters resulting 
from the quantum perturbational approach and related algebraic representations (see e.g.
\cite{rudowicz_concept_1987,*rudowicz_erratum_1988} and references therein), we need only the spin Hamiltonian
\begin{equation}\label{eq:sds_hamilton}
	\hat{\mathcal{H}}=\hat{\mathbf{S}}\cdot\mathbf{D}\cdot\hat{\mathbf{S}},
\end{equation}
where $\hat{\mathbf{S}}=(\hat{S}_\alpha)_{\alpha\in\mathbb{K}}$ is the effective spin-one operator of the considered system, 
$\mathbf{D}$ is a symmetric tensor, with elements $D_{\alpha\beta}\in\mathbb{R}$, $\alpha,\beta\in\mathbb{K}$.
Hereafter, we will see that a direct mapping between the spectrum of \eqref{eq:sds_hamilton} and 
those depicted on Fig. \ref{fig:en} (b) is not feasible.

With respect to the applied method, the FS relevant to the low-lying energy levels 
is quantified by the energy gaps 
$\Delta E_{21}=0.1708$, $\Delta E_{31}=0.3649$ and $\Delta E_{32}=0.1941$ meV. 
The associated functions are the ground state \eqref{eq:ZGstate}, first \eqref{eq:ZExstate} 
and second \eqref{eq:Ap_ZExstate_2} 
excited states. Represented only within the effective spin space they 
read $|\Psi_1\rangle\to C(|0\rangle-|\bar{1}\rangle)$, 
$|\Psi_2\rangle\to|1\rangle$ and $|\Psi_3\rangle\to C(|0\rangle+|\bar{1}\rangle)$, 
respectively, where $C=${\small $1/\sqrt{2}$}.
Here, the superposition of the ground and second excited states cannot be addressed 
by the spin Hamiltonian \eqref{eq:sds_hamilton} since in our case the
CF contributes significantly 
to the mixing of states.
As a consequence, we could perform a sort of mapping and extract useful information for the elements of $\mathbf{D}$ only 
by rising the complex's symmetry to $C_{3v}$ and preserving the overall ZFS that equals $0.3649$ meV.
Now, calculating the energy level sequence and associated eigenstates
in the case of ideal trigonal 
bipyramidal geometry, we have $D_{\alpha\beta}=0$ for all $\alpha\ne\beta$, $D_{yy}=D_{xx}=\Delta E_{31}/3$ 
and $D_{zz}=-2\Delta E_{31}/3$ meV. 
Owing to the relation between ZFS parameters $(D, E)$ and the tensor elements 
\protect\cite{abragam_electron_2012,kozanecki_high-frequency_2017}, 
we get $D=-\Delta E_{31}$ and $E=0$.

In contrast to the positive value of $D$ obtained in Ref. \cite{fataftah_trigonal_2020} and hence $|0\rangle$ 
spin-one ground state, our calculations suggests the opposite scenario. 
In other words, the ground state energy level should be $2$-fold
degenerate and represented by the
$|1\rangle$, $|\bar{1}\rangle$ states, with $D\approx-2.943$ cm$^{-1}$ and $|0\rangle$ being the excited state.


\section{Magnetic and spectroscopic properties}\label{sec:magsub}

\subsection{Magnetization and susceptibility}

\begin{figure}[h!]
	\centering
	\includegraphics[scale=1.2]{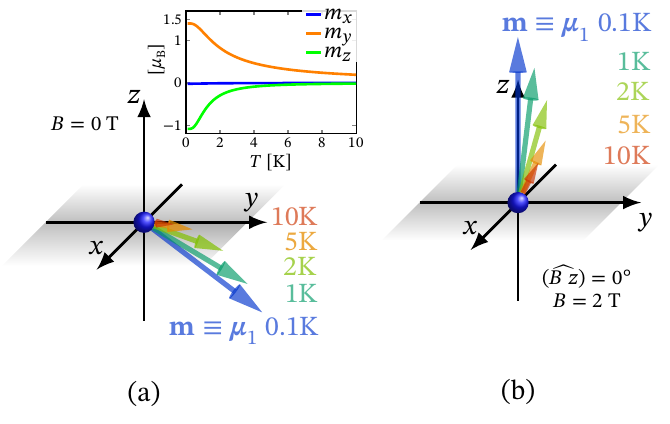}
	\caption{Evolution of the  
		effective magnetic 
		moment \eqref{eq:Magm} in the temperature domain $0.1\le T\le10$ K. 
		The blue ball represents the
		vanadium ion residing at the center of the local
		reference frame with orientation depicted
		on Fig. \ref{fig:stucture}. 
		(a) Shows the temperature dependence in the absence of
		an external magnetic field.
		(b) Depicts the case of an applied along the $z$ axis magnetic field, 
		with magnitude $B=2$ T.
	}
	\label{fig:MagMomT}
\end{figure}

\begin{figure}[ht!]
	\centering
	\includegraphics[scale=1]{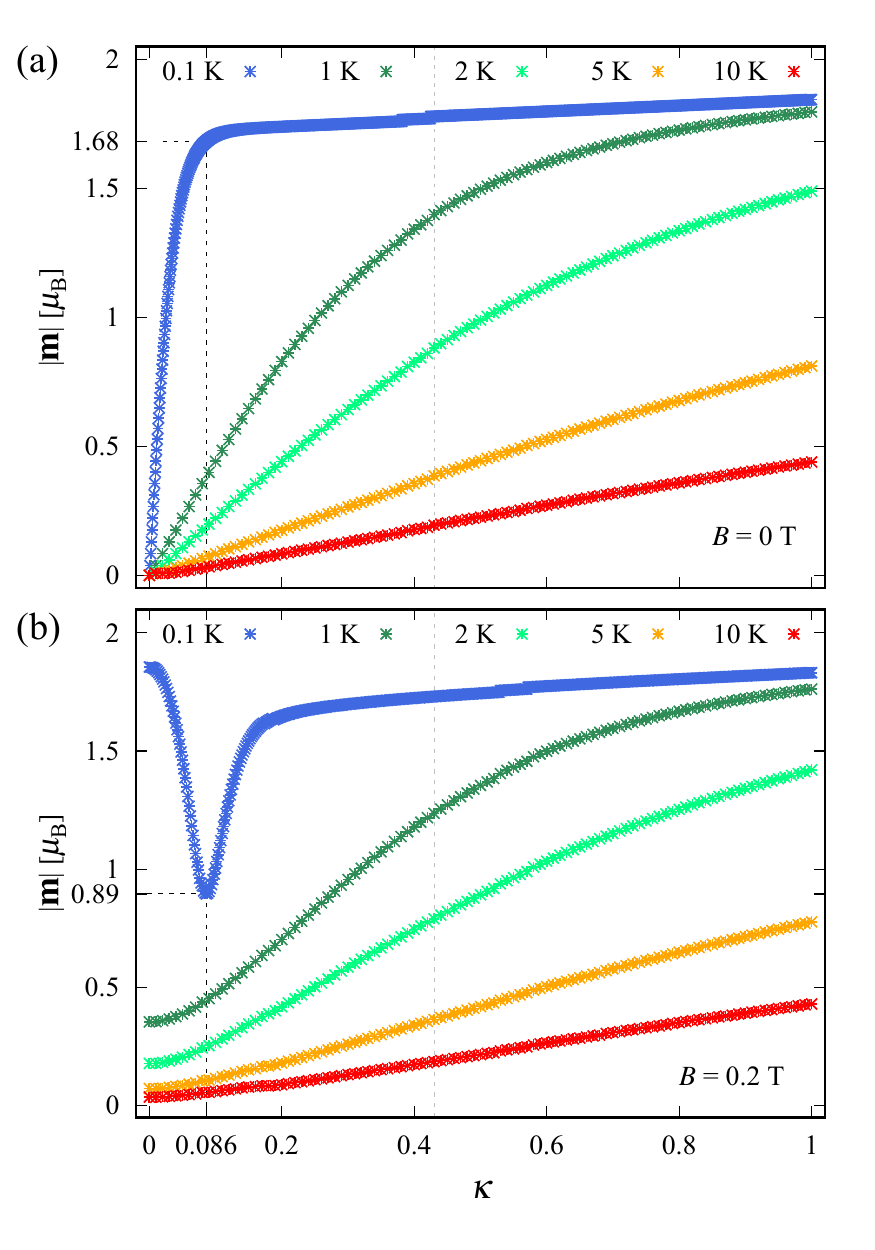}
	\caption{The magnitude of magnetic moment \eqref{eq:Magm} as a function of reduction coefficient $\kappa$ 
			at selected temperatures. 
		The dashed gray line shows the value of $\kappa$ obtained from the comparison with the experimental data from 
		Ref. \cite{fataftah_trigonal_2020} and given in Tab. \ref{tab:strpar}.
		(a) Shows the $\kappa$ dependence in the absence of the external magnetic field.
		(b) Depicts the low-field case, with $B\equiv B_z=0.2$ T. The dashed black line indicates the minimum of 
		$|\mathbf{m}|$ at $T=0.1$ K 
		due to the interplay between Zeeman and spin-orbit interactions.
	}
	\label{fig:Mzfs}
\end{figure}

The obtained FS in the zero-field energy spectrum 
shown on Fig. \ref{fig:en} (b), with energy gaps smaller than half of a meV, 
clearly indicates that the compound's magnetic moment 
\begin{equation}\label{eq:Magm}
\mathbf{m}(T)=\varOmega^{-1}\sum^{45}_{n=1}\boldsymbol{\mu}_n e^{-E_n/k_B T},
	\ \text{with} \ \
	\varOmega=\sum_{n=1}^{45}e^{-E_n/k_B T},
\end{equation}
rapidly decreases by magnitude with increasing temperature.
The dependance on $\kappa$ and therefore on the overall ZFS, is shown on Fig. \ref{fig:Mzfs} (a), 
where $0\leq\kappa\leq1$. 
Thus, for $\kappa\le0.1$ the effect of temperature is most prominent in the domain $0.1$ to $1$ K.
On the contrary, when the ZFS is large, or $\kappa=1$, in the same temperature domain the magnitude 
of $\mathbf{m}$ remains almost unchanged.
Moreover, as it is depicted on Fig. \ref{fig:MagMomT} (a), $\mathbf{m}$ 
also deviates from its initial direction when $T\to0$.

Even at low magnetic field values, the small ZFS renders the magnetic behavior of the considered complex very susceptible 
to temperature variations, see Fig. \ref{fig:Mzfs} (b). 
An exception is the domain $0\le\kappa\le0.086$ for $T\le0.1$ K, where
the contribution of the externally applied 
magnetic field to the mixing of the ground state is greater than the spin-orbit one.
Hence, the smaller the ZFS the greater the population \cite{kittel_2005}
of first two excited states for a given temperature.
As an example, for $\kappa=0.43$, $T=10$ K and $B=0.1$ T the population rates of the
first two excited 
energy levels are almost equal. 
In particular, we have the probability to observe 
$40\%$ of the constituents in the powder sample at the ground 
state, $p_2\approx33.5\%$ in the first excited state and 
$p_3\approx26.2\%$ in the third one.
As a result, we obtain a prompt decrease in the low-field
susceptibility in
the temperature domain $0<T\le10$ K. A comparison between theory and experiment is 
depicted on Fig. \ref{fig:TST}, where the fraction between the corresponding molar 
mass and density is found to be approximately $0.975$ [cm$^3$/mol]. 

Since the zero-field spectrum is non-degenerate the action of
the externally applied 
magnetic field changes the FS of the energy spectrum by only 
increasing the corresponding energy gaps, see for example Fig. \ref{fig:en} (c). 
The eigenstates also change, such that their magnetic sequence resembles the spin
Zeeman splitting.  
Thus, while reaching saturation of the magnetization, the probability 
of observing a fully polarized magnetic moment, along the
magnetic field direction, at 
the ground state equals unity. 

With magnetization data collected on a powder sample 
the magnetization $\mathbf{M}(B, T)$ is 
computed as an arithmetic mean of the magnetic moments $\mathbf{m}$ of all 
complexes that can be represented as a parity transformation one to another 
regarding their 
local reference frame shown on Fig. \ref{fig:stucture}. 
The temperature dependence of magnetization per complex, $M=|\mathbf{M}(T)|$,
for some external magnetic field values  
is depicted on Fig. \ref{fig:Mag}. The consistency with the
experimental data from Ref. \cite{fataftah_trigonal_2020} is obtained for  
values of the model parameters in Tab.
\ref{tab:strpar} and the angle $\nu$
between $\mathbf{M}(B, T)$ and $\mathbf{B}$ in the
bottom row of Tab. \ref{tab:Magfield}. 

\begin{figure}[ht!]
	\centering
	\includegraphics[scale=1]{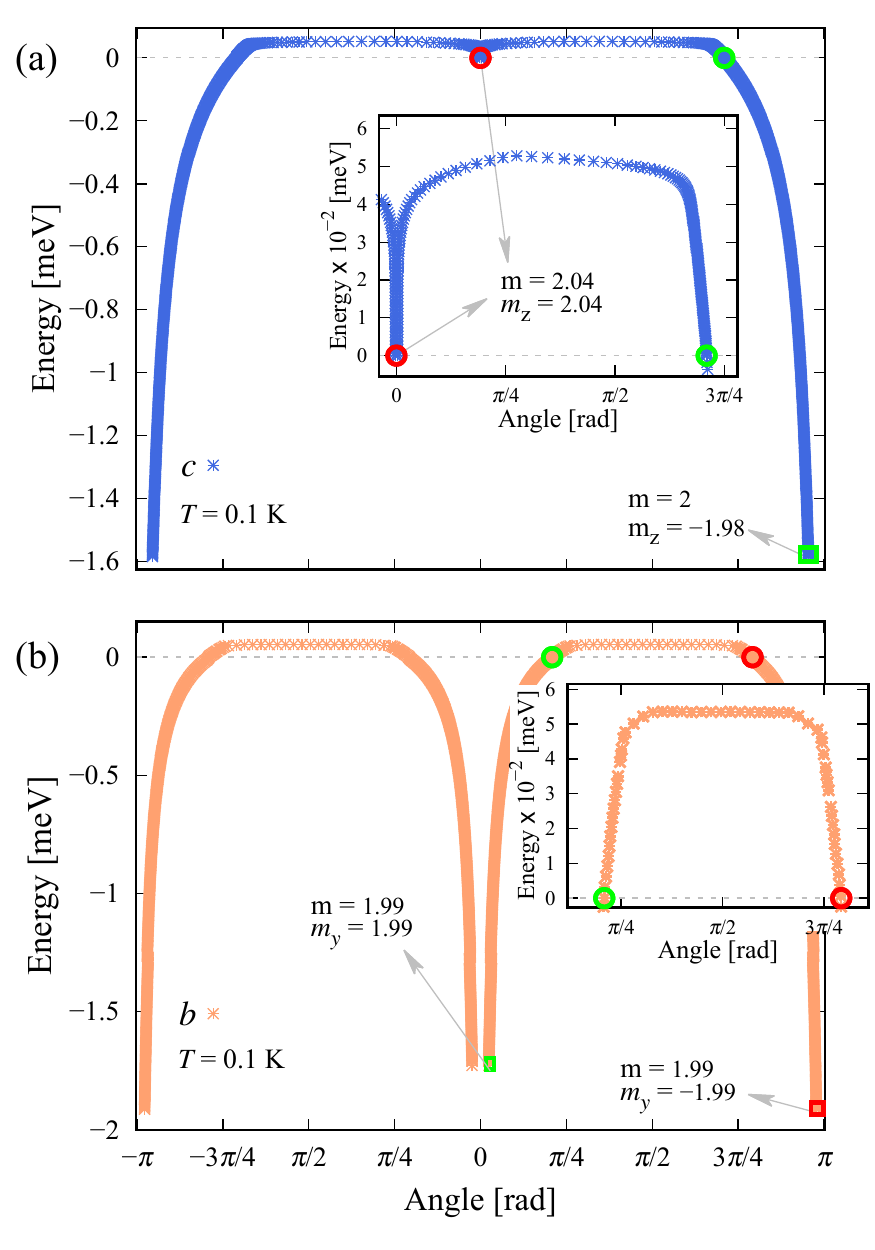}
	\caption{Ground state energy of a single molecule as a
	function of the direction angles of the magnetic 
			moment \eqref{eq:Magm} at $T=0.1$ K and parameters given in Tab. \ref{tab:strpar}. 
			The energy is normalized with respect to its minimum at 
			$B=0$, indicated by a green circle and with magnetic moment components shown on Fig. \ref{fig:MagMomT} (a).
			The second energy minimum that could be equally occupied is indicated by a red circle.
			Both insets show the energy barrier profile and its center between the two zero-field minima.
			The minimal energy to both fully polarized states $m_\alpha\to\pm2$, with $\alpha=z,y$, is indicated by squares.
			(a) depicts the dependence on the angle ``$c$'' between 
			$\mathbf{m}$ and the $z$ axis of the compounds reference frame, see Fig. \ref{fig:stucture}, and  
			(b) shows the ground state energy as a
			function of the angle ``$b$'' between $\mathbf{m}$ and the 
			$y$ axis of the compounds reference frame.
		Notice that no energy barrier is observed along the $x$ axis.
	}
	\label{fig:ZYans}
\end{figure}

\begin{figure}[ht!]
	\centering
	\includegraphics[scale=1]{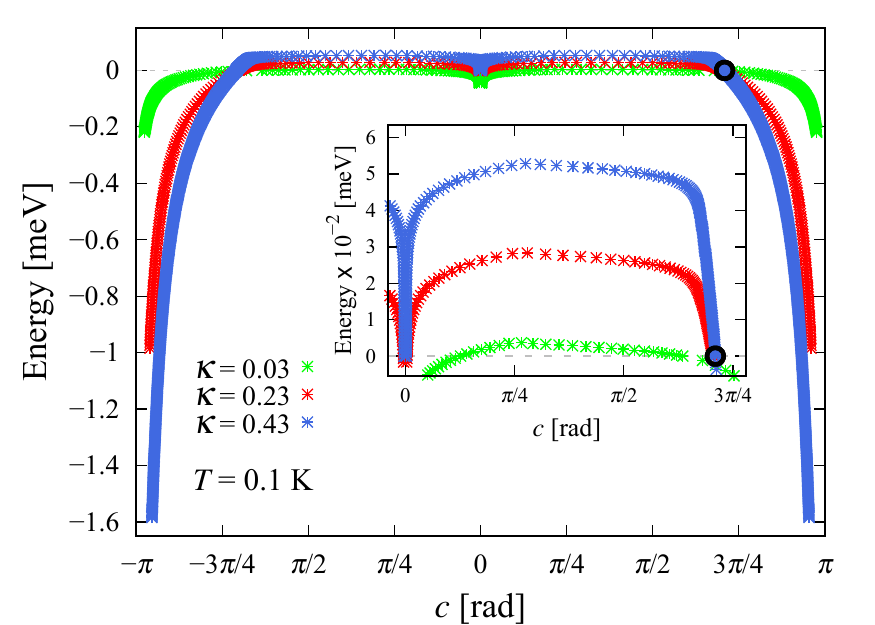}
	\caption{The ground state energy of a single molecule as a function of the direction angle ``$c$'' between 
			$\mathbf{m}$ and the $z$ axis of the compounds reference frame and its dependence on the  
			reduction parameter $\kappa$ to the spin-orbit
			coupling. No angle dependence is observed at $\kappa=0$.
			The energy is normalized to the value obtained in the case $B=0$, indicated with a black circle.
			For additional details see Fig. \ref{fig:ZYans} (a).
			The inset depicts detailed view of the energy barrier profile.
	}
	\label{fig:Zk}
\end{figure}

\begin{figure}[ht!]
	\centering
	\includegraphics[scale=1]{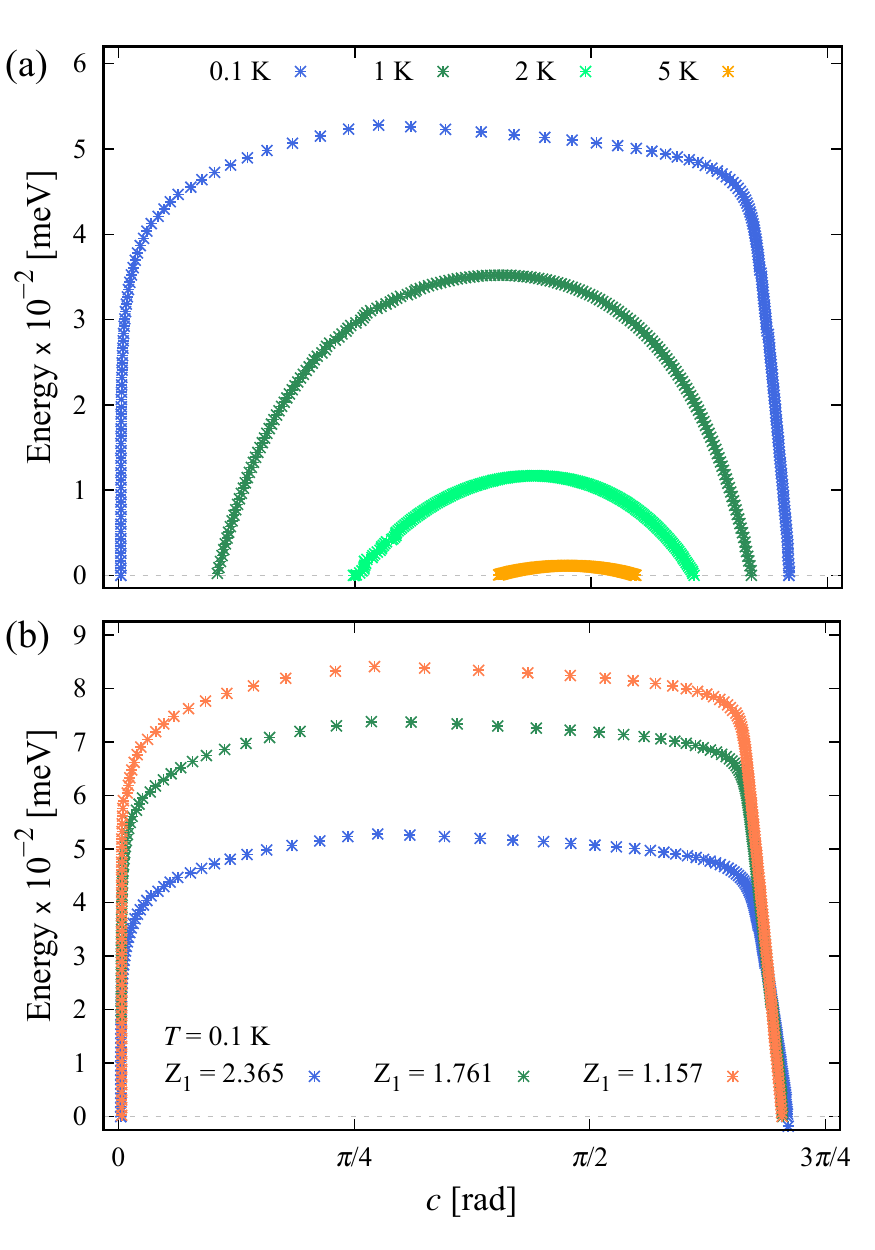}
	\caption{(a) Temperature dependence of the energy barrier calculated with respect to the $z$ axis. 
			The excited states contribute for $T>0.1$ K. 
			More details are shown on the inset in Fig. \ref{fig:ZYans} (a).
			Subfigure (b) depicts how the barrier's height along the $z$ axis changes with respect to the 
			cyanide charge number $Z_1$ at $T=0.1$ K.
	}
	\label{fig:TCN}
\end{figure}

At $7$ T and $T\le1$ K the magnetization is 
approximately $0.963$ times its maximal value reached at about $B=10$ T.
In the case of low field values the obtained small reduction of the magnetization at $B=1$ T and
$T\le0.8$ K, shown on Fig. \ref{fig:Mag}, is due to the 
presence of single-ion anisotropy, see the energy barriers depicted on Figs. \ref{fig:ZYans} and \ref{fig:Zk}.
As an example, taking $B=1$ T and $T=0.8$ K, for the magnetization of a powder sample we have $M=0.6413$.
Yet calculating the magnetic moment \eqref{eq:Magm}, for   
{\small $\{\widehat{\mathbf{B} \ \pm x},\widehat{\mathbf{B} \ \pm y},\widehat{\mathbf{B} \ z}\} =$}
$\{58.53^{\circ},58.53^{\circ},47.46^{\circ}\}$, 
we get $\mathbf{m}_{+}=(0.582,1.455,-0.215)$ and $\mathbf{m}_{-}=(-0.739,0.195,0.466)$. 
Hence, the energy required to fully polarize a single molecule strongly depends on the direction 
of the externally applied magnetic field in the compound's reference frame. 
In order to shed more light on the existing single-ion anisotropy, we explore how the complex's energy 
depends on the direction angles $\{a,b,c\}$ \cite{kittel_2005,abragam_electron_2012,weber_essentials_2014} of the magnetic moment \eqref{eq:Magm} 
taken with respect to the reference frame shown on Fig. \ref{fig:stucture}.
Since the $z$ axis coincides with the principle axis of the ideal trigonal geometry case and we do not 
have a $\bot C_2$ axis of symmetry, we set 
$\{a,b,c\}=\{\widehat{\mathbf{m} \ x},\widehat{\mathbf{m} \ y},\widehat{\mathbf{m} \ z}\}$.
Moreover, in view of the fact that at $T=0.1$ K the ground state is about a hundred percent populated, 
excited energy levels have a negligible contribution to the obtained energy barrier profile.
Therefore, we obtain a small energy barrier with height
$E_{br}\approx\Delta E_{21}/3$ meV, shown on the insets in Fig. \ref{fig:ZYans} and \ref{fig:Zk}.
The barrier separates two different states of equal population in the case $T=0.1$ K, one of which is the ground state 
\eqref{eq:ZGstate} indicated by a green circle. 
The representation of the second state, shown by a red circle, depends
on the axis along which the magnetic moment
$\mathbf{m}$ reverses. For example, considering the $z$ axis, with barrier depicted on Fig. \ref{fig:ZYans} (a), 
we find that the second state is \eqref{eq:ZExstate}. 
The evolution of the energy barrier with respect to $B$ is depicted on the background in Figs. \ref{fig:ZYans} and \ref{fig:Zk}.
The value of $E_{br}$ depends on the direction of the applied magnetic
field.
Taking the $z$ axis, for example, with $\mathbf{B}=(0,0,B_z)$, we get $E_{br}\approx\Delta E_{21}/3$ meV.
On the other hand, for $\mathbf{B}=(0,0,-B_z)$ the barrier's height is approximately $1.6$ meV.
The most prominent value of $E_{br}$ is obtained in the case $\mathbf{B}=(0,-B_y,0)$, reaching nearly $2$ meV.
The energy barrier rapidly vanishes by increasing the temperature. 
The corresponding dependence is shown on Fig. \ref{fig:TCN} (a) and is  
intrinsically related to the evolution of $\mathbf{m}$ depicted on Fig. \ref{fig:MagMomT} (a).
Moreover, Fig. \ref{fig:Zk} shows that the barrier's height reduces significantly by decreasing the value of $\kappa$, 
demonstrating how it depends on the ZFS sketched on Fig. \ref{fig:en} (b).
Note that as expected no energy barrier is observed along the $x$
axis, by virtue of the temperature dependence of
$m_x$ shown on Fig. \ref{fig:MagMomT} (a). 

\subsection{cw-EPR}

According to the cw-EPR experimental observations reported in Ref. \cite{fataftah_trigonal_2020} 
the Vanadium based complex responds to a $240$ GHz microwave radiation at approximately 
$3.75$, $6.55$ and $10.4$ T magnetic field values and $T=8$ K. 
Since all three transitions are related to a spin triplet states the selection rules 
are $\Delta s=0$ and $|\Delta m|=1,2$. 
We would like to point out that
these transitions are selected for the sake of convenience  
and for mixed states the associated selection rules should be
interpreted in accordance with Refs. \cite{abragam_electron_2012,rudowicz_selection_2021}. 
The corresponding excitations, with energy approximately $0.992$ meV, 
are reasonably well reproduced by the present theoretical approach.  
They are related to the existing anisotropy that
determines the $y$ 
component of the magnetic moment
\eqref{eq:Magm} as a dominant transverse component in the whole temperature range, 
see Fig. \ref{fig:MagMomT} (a) and the energy barriers depicted on Fig. \ref{fig:ZYans}. 
In other words, with respect to the compounds reference frame illustrated 
on Fig. \ref{fig:stucture}, the resonance condition is satisfied only when the magnetic field vector  
lies in the $(zy)$ plane and forms a certain angle with the $z$ and $y$ axes.
Thus, only certain units in the powder sample with the
appropriate orientation with respect to
the direction of $\mathbf{B}$ will be magnetically excited, contributing on average to the change 
in magnetization from the perspective of laboratory reference frame.

By taking $\mathbf{B}=B\mathbf{n}$, where $\mathbf{n}=(n_\alpha)_{\alpha\in\mathbb{K}}$ 
is the field's unit vector with respect to the compound's reference frame, 
we obtain that the resonance occurs for $n_x=0$, $0.4212\le n_y\le 0.4338$ and 
$-0.901\le n_z\le-0.907$.

The first transition at $3.75$ T refers to a magnetic excitation with $|\Delta m|=2$. 
It involves the ground and the second excited states.
The population rate at the ground state is $62.5\%$ and 
represented by a superposition of the three spin-one triplets 
with coefficients given in percentage, it reads 
$|\Psi^{3.75}_1\rangle\to0.1\%|1,1\rangle+23.8\%|1,0\rangle+76.1\%|1,\bar{1}\rangle$.
Respectively, for the second excited state, we get 
$|\Psi^{3.75}_3\rangle\to81.3\%|1,1\rangle+16.3\%|1,0\rangle+2.4\%|1,\bar{1}\rangle$.

The second excitation at $6.55$ T is associated to a transition
between the ground and the first 
excited states, with $|\Delta m|=1$. The calculated population
of the ground state is about $75.2\%$,
which explains the higher intensity of the relevant peak
compared to that of the first excitation.
With respect to the probabilities distribution among the three spin triplets, 
both states read
$|\Psi^{6.55}_1\rangle\to1.7\%|1,1\rangle+18.9\%|1,0\rangle+80.4\%|1,\bar{1}\rangle$ and
$|\Psi^{6.55}_2\rangle\to13.2\%|1,1\rangle+68.4\%|1,0\rangle+18.4\%|1,\bar{1}\rangle$.

The third peak at approximately $10.4$ T is associated to a transition between the first and second 
excited states that can be represented as
$|\Psi^{10.4}_2\rangle\to11.9\%|1,1\rangle+72.4\%|1,0\rangle+15.7\%|1,\bar{1}\rangle$ and
$|\Psi^{10.4}_3\rangle\to88.5\%|1,1\rangle+11\%|1,0\rangle+0.5\%|1,\bar{1}\rangle$, respectively.
The corresponding selection rule is $|\Delta m|=1$.
The population of the first excited level at $8$ K is about $10.1\%$ which is consistent with 
the observed low intensity of the peak \cite{fataftah_trigonal_2020}.

\subsection{Photoluminescence}

The photoluminescence spectroscopy performed on (C$_6$F$_5$)$_3$trenVCN$^t$Bu in 
2-methyltetrahydrofuran at $77$ K, 
reported in Ref. \cite{fataftah_trigonal_2020},
shows a continuous near-infrared emission with spectral peak at approximately $1.240$ $\mu$m
in the excitation wavelength range $550-750$ nm. 
According to the $s=1$ ground state, the transitions 
to all photo-excitations are triplet-singlet related. Respectively, the observed
phosphorescence is associated to a singlet-triplet relaxation process.
Furthermore, the large difference between the excitation and emission energies
suggests that the overall relaxation pathway
may incorporate
transitions with different vibronic modes.
To this end, in an attempt to reproduce the photoluminescence data, we simulate 
variations in the polar and azimuthal angles for all ligands with
a value no larger than $0.63^\circ$, since within the domain $(-0.63^\circ, 0.63^\circ)$
the ZFS depicted on Fig. \ref{fig:en} (b) remains invariant. 
This simulation allows us to trace the corresponding change of the 
electronic energy levels and the extent of spin-multiplet mixing in the
resulting eigenstates. 
Actually, the phonon energies and interactions are not considered. 
Moreover, for brevity variations in the radial position of the 
ligands are not computed.

For $\Delta\vartheta_i=0^\circ$ and $\Delta\varphi_i=0^\circ$, $i=1,\ldots,5$,
the function \eqref{eq:S26} is the first singlet state with 
energy $E_{26}=1.84694$ eV, see App. \ref{app:enspectrum}. At this 
point the group of eigenstates 
$\{|\Psi_{23}\rangle,|\Psi_{24}\rangle,|\Psi_{25}\rangle\}$ are all $s=1$ states 
with more than $99.9\%$ probability.
For example
\begin{align*}
|\Psi_{23}\rangle=&
0.55i |\psi_{10,1,0}\rangle - 0.38|\psi_{10,1,\bar{1}}\rangle - 0.38|\psi_{10,1,1}\rangle 
	\\
& \! + 0.45i|\psi_{9,1,0}\rangle - 0.31|\psi_{9,1,\bar{1}}\rangle - 0.31|\psi_{9,1,1}\rangle
	\\ &
- 0.08i|\psi_{7,1,0}\rangle + 0.055|\psi_{7,1,\bar{1}}\rangle + 0.055|\psi_{7,1,1}\rangle
	\\ &
+0.048i|\psi_{4,1,0}\rangle - 0.033|\psi_{4,1,\bar{1}}\rangle - 0.033|\psi_{4,1,1}\rangle
	\\ &
	+ \sum_{n\ge2} O\big(10^{-n}\big)|\psi_{\ldots}\rangle.
\end{align*}
In this case, the lower possible excitation wavelength 
lies in the experimentally observed domain 
and is approximately $671$ nm. However, the only
spin-allowed  emission near to 
the observed one, with no intersystem crossing, is of low probability and related to the transition 
$|\Psi_{26}\rangle\to|\Psi_{15}\rangle$. It is about $1.16$ $\mu$m, where
\begin{align}\label{eq:S15}
	|\Psi_{15}\rangle=&
	(-0.61-0.27i)|\psi_{1,1,0}\rangle + (0.26-0.6i)|\psi_{1,1,\bar{1}}\rangle 
\nonumber	\\ &
	- (0.17+0.08i)|\psi_{2,1,0}\rangle + (0.07-0.17i)|\psi_{2,1,\bar{1}}\rangle 
\nonumber	\\ &
	+ (0.039+ 0.018i)|\psi_{5,1,0}\rangle - (0.017-0.039i)|\psi_{5,1,\bar{1}}\rangle 
\nonumber	\\ &
	- (0.1+0.044i)|\psi_{6,1,0}\rangle + (0.042-0.099i)|\psi_{6,1,\bar{1}}\rangle 
\nonumber	\\ &
	+ (0.113+0.051i)|\psi_{8,1,0}\rangle - (0.048-0.112i)|\psi_{8,1,\bar{1}}\rangle 
\nonumber	\\ &
	+ \sum_{n\ge2} O\big(10^{-n}\big)|\psi_{\ldots}\rangle.
\end{align}
\begin{figure}[th!]
	\centering
	\includegraphics[scale=1]{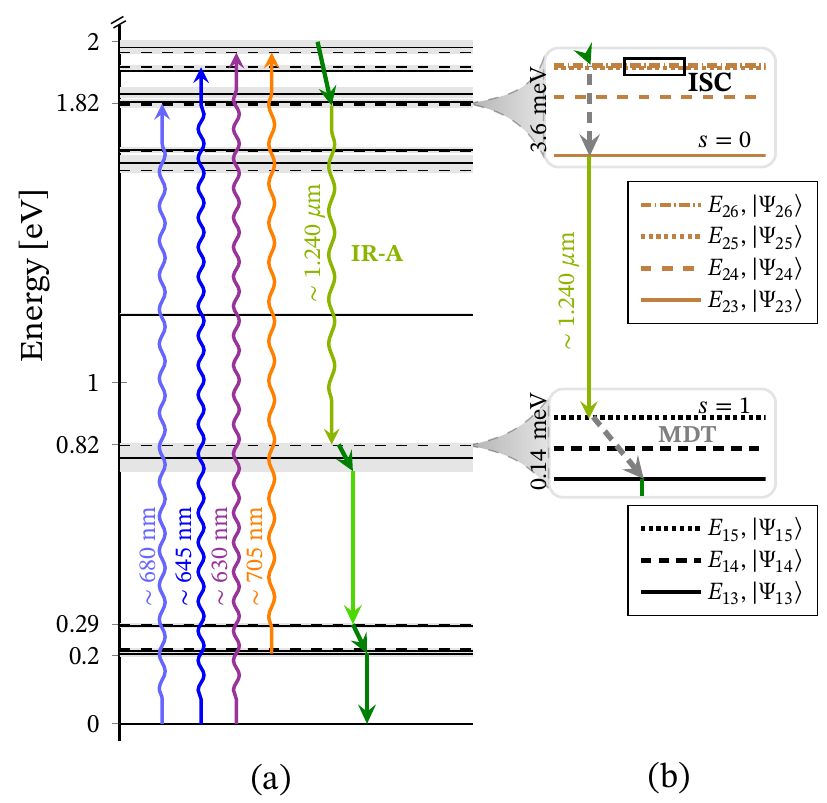}
	\caption{(a) Energy level diagram of (C$_6$F$_5$)$_3$trenVCN$^t$Bu 
		up to $2$ eV with possible absorption bands and relaxation pathway 
		reproducing the photoluminescence and UV--Vis--NIR spectroscopy
		measurments reported in Ref. \cite{fataftah_trigonal_2020}. 
		For all $i=1,\ldots,5$ the black lines represent the case with
		$\Delta\vartheta_i=0^\circ$, $\Delta\varphi_i=0^\circ$ shown on Fig \ref{fig:en} (a). 
		The gray solid fields depict bands of energy levels associated to 
		a possible continuous or discrete variations in the angular 
		positions of all ligands, in the intervals 
		$-0.63^\circ\le\Delta\vartheta_i\le0.63^\circ$, 
		$-0.63^\circ\le\Delta\varphi_i\le0.63^\circ$.
		The blue, violet and orange arrows refer to
		spin-allowed
		ground and excited state transitions, respectively, where
		$\Delta\vartheta_i=0.585^\circ$, $\Delta\varphi_i=0^\circ$.
		Accordingly, the olive green arrow depicts singlet-triplet radiative 
		decay at $1240$ nm related to the observed emission, see 
		Ref. \cite{fataftah_trigonal_2020}. 
		Non-radiative decays are illustrated by solid non-curved arrows. 
		The light-green arrow shows a transition at about $3$ $\mu$m.
		(b) A magnification shows the FS for the case $\Delta\vartheta_i=0.585^\circ$, 
		$\Delta\varphi_i=0^\circ$, 
		with possible intersystem crossing (ISC) and magnetic dipole 
		transitions (MDT) depicted by dashed gray arrow.  
	}
	\label{fig:enpl}
\end{figure}

A better theoretical description of the photoluminescence spectroscopy is obtained 
by varying the polar and azimuthal angles of all ligands with an amount of 
$\pm0.63^\circ$ to their initial values given in Tab. \ref{tab:strpar}.
The resulting emission wavelengths take values in the
range $1.230-1.250$ $\mu$m.
A particular case with $\Delta\vartheta_i=0.585^\circ$ and 
$\Delta\varphi_i=0^\circ$ for all $i=1,\ldots,5$ is depicted on Fig. 
\ref{fig:enpl} (a).
The most considerable is the shifting related to the observed emission involving 
the set of energy levels $\{E_{13},E_{14},E_{15}\}$.
An absorption band and a possible relaxation pathway are depicted by curved 
arrows, where the absorption at $705$ nm is of very low probability rate for 
$T<300$ K. The corresponding radiative decay at about $1240$ nm is a 
singlet-triplet transition.
In particular, the compound undergoes a transition from $|\Psi^{\Delta}_{23}\rangle$ to 
$|\Psi^{\Delta}_{15}\rangle$, where the former state is now
the $99.5\%$ singlet 
\begin{align*}
	|\Psi^{\Delta}_{23}\rangle=&
-0.11|\psi_{1,0,0}\rangle + 0.25|\psi_{2,0,0}\rangle - 0.2|\psi_{3,0,0}\rangle 
\\ &
- 0.73|\psi_{5,0,0}\rangle - 0.016|\psi_{6,0,0}\rangle - 0.031|\psi_{7,0,0}\rangle 
\\ &
+ 0.25|\psi_{8,0,0}\rangle + 0.045|\psi_{9,0,0}\rangle - 0.051|\psi_{11,0,0}\rangle   
\\ &
+ 0.26|\psi_{12,0,0}\rangle + 0.38|\psi_{13,0,0}\rangle - 0.13|\psi_{14,0,0}\rangle
\\ &
+ \sum_{n\ge1} 
O\big(10^{-n}\big)|\psi_{\mathrm{trip}.}\rangle.
\end{align*}
The superposition in the energy eigenstate
$|\Psi^{\Delta}_{15}\rangle$ has almost the same expression
as \eqref{eq:S15}.
For more details see App \ref{app:photo}. 
In addition, Fig. \ref{fig:enpl} (b) depicts the FS
of this transition. 
Depending on the values of $\Delta\vartheta_i$ and $\Delta\varphi_i$, 
the calculations predict a different extent of intersystem crossing 
between two group of levels related to the states 
$\{|\Psi^{\Delta}_{23}\rangle,|\Psi^{\Delta}_{24}\rangle,|\Psi^{\Delta}_{25}\rangle\}$
and 
$\{|\Psi^{\Delta}_{26}\rangle,|\Psi^{\Delta}_{27}\rangle,|\Psi^{\Delta}_{28}\rangle\}$.
The relaxation to the ground state after the emission
involves only spin triplet
states and is assumed to be related to vibrational modes. 

In spite of the fact that the
proposed approach sheds light on
the experimentally observed phosphorescence,
the decay corresponding to a wavelength of nearly $3\mu$m remains
unresolved (see Fig. \ref{fig:enpl} (a)).

We would like to point out that the isocyanide member may allow significant 
stretching of its bond to the vanadium center, as it is discussed 
in Ref. \cite{fataftah_trigonal_2020}.
In this respect, variations equal or less than $\pm0.05$ \AA{} in the 
corresponding radial position 
has no significant contribution to the energy 
spectrum shown on Fig. \ref{fig:enpl} (a). Yet, any value closer to $\pm0.1$ 
\AA{} considerably increases the shifting of the energy levels and leads to 
a different ZFS.


\section{Summary}\label{sec:Conclusion}

To gain useful insight into the magneto-structural dependencies in $3d^2$ systems and the 
mutual influence of crystal field, exchange and spin-orbit
interactions on their magnetic behavior, 
we performed an extensive and detailed theoretical investigation of the magnetic and 
spectroscopic properties of the compound (C$_6$F$_5$)$_3$trenVCN$^t$Bu. 
Using a multi-configurational method with active space restricted to all 
five $d$ orbitals, 
we reproduce and interpret the available in literature experimental data 
\cite{fataftah_trigonal_2020}. 
This includes calculations for the magnetization, low-field susceptibility, 
cw-EPR and photoluminescence spectroscopy. For more details see e.g. Sec. \ref{sec:magsub}. 

In general, in the parameterization scheme associated to the
proposed
method there are three sets of non-variational parameters.
The first set includes the coordinates of all ligands, obtained from the structural analysis. 
The second comprises of the charge numbers of all ligands and metal centers, see Sec. 
\ref{sec:Hamilton}, in particular \eqref{eq:CF_term_1}. 
Moreover, the second set accounts for a control parameter to the spin-orbit coupling, 
denoted in the present study by $\kappa$ and discussed at the beginning of Sec. \ref{sec:SO}.
Finally, the third contains the components of the unit vector associated to the 
applied magnetic field, taken with respect to the compound's
reference frame (see e.g. Fig. \ref{fig:stucture}). 
It further includes all derived quantities, such as the effective
angle given
in Tab. \ref{tab:Magfield}.
Let us stress that these parameters uniquely characterize the compound under consideration.

In particular, the study of (C$_6$F$_5$)$_3$trenVCN$^t$Bu points out that the
electrons are partially delocalized. This may be traced back to the 
lower than expected values of the prefactor $\kappa$ and 
charge number $Z$ of the vanadium ion, see Tab. \ref{tab:strpar}.
This is also evident from the larger than expected value of $Z_1$.
Furthermore, 
the calculations show that the difference between the values of
the charge number of nitrogen and isocyanide ligands
plays a significant role in shaping the low-field low-temperature
magnetic properties.
Accordingly, there is an optimal range for this difference where
one obtains good agreement with the available experimental data.
For example, reducing the value of $Z_1$ from Tab. \ref{tab:strpar}  
results in a better agreement with the low-susceptibility measurements and 
in a worse reproduction of the low-field magnetization, cw-EPR and photoluminescence data. 
This is due to the corresponding decreasing in energy of the $d_{xz}$ and $d_{yz}$ orbitals and increasing of ZFS. 
However, the value of $Z_1$ does not contribute considerably to the 
zero-field ground state magnetic properties. In particular, setting $Z_1=Z_i$, for all $i=2,\ldots,5$, with 
$Z_i$ given in Tab. \ref{tab:strpar}, the orientation of the total magnetic moment at the 
ground state slightly changes from $\boldsymbol{\mu}_1=(-0.0171, 1.4091, -1.0865)$
to $\boldsymbol{\mu}_1=(-0.02695, 1.36805, -1.01942)$ (see Fig. 
\ref{fig:MagMomT} (a) and Sec. \ref{sec:zfs}). 
As a result, for $T\to0$ the magnitude and orientation of $\mathbf{m}$ given in \eqref{eq:Magm} slightly change.
The $d_{xz}$, $d_{yz}$ orbitals are now lower in energy than the $d_{xy}$, $d_{x^{2}-y^{2}}$ ones
and the ZFS is stronger. We have $\Delta E_{21}=0.2855$, $\Delta E_{31}=0.5785$ and $\Delta E_{32}=0.2929$ meV.
Consequently, the population of the corresponding levels changes
leading to a different temperature dependence of 
the magnetic moment \eqref{eq:Magm} in the case $T>0$.
Furthermore, the anisotropy energy \cite{kittel_1987,kittel_2005,jensen_rare_1991,buschow_physics_2003} 
also changes, such that the corresponding energy barrier's height increases. 
An example is depicted on Fig. \ref{fig:TCN} (b), where 
for $T=0.1$ K and $Z_1=1.157$ with respect to the $z$ axis we have $E_{br}\approx0.084$ meV. 

Let us stress that in the case of ideal trigonal bipyramidal geometry
we obtain an energy level sequence that can be directly related to the conventional ZFS Hamiltonian \cite{rudowicz_concept_1987,*rudowicz_erratum_1988,rudowicz_spin-hamiltonian_2001}. 
Thus, one has the case of $2$-fold degenerate ground state 
and an excited state related to $m=0$ spin state, with $D<0$ and $E\to0$. 
Nevertheless, the proposed multi-configurational scheme still gives the energy eigenstates
as a linear combination of different configuration functions.
The comparison with the axial $D$ and rhombic $E$ ZFS parameters is provided in Sec. \ref{sec:DE}.

An essential feature observed in the presence of external magnetic field is the evolution of FS.
The shifting of energy levels due to the action of $\mathbf{B}$ is discussed in Sec. \ref{sec:MagFieldEffect}.
Nevertheless, another example of how the Zeeman and spin-orbit terms interfere causing this effect 
is evident from the dependence of magnetic moment \eqref{eq:Magm}  
on the reduction coefficient $\kappa$. This effect is prominent for low-temperatures and is demonstrated on
Fig. \ref{fig:Mzfs} (b). Within the studied low symmetry complex, 
there exist a minimum value of $|\mathbf{m}|$ that corresponds to a maximal degeneracy
of the ground state energy level in the case $B\ne0$. With respect to the case depicted on Fig. \ref{fig:Mzfs} (b),
for $\kappa=0.086$ and $B\equiv B_z=0.2$ T, the ground state level is $2$-fold degenerate. 
As a result, the magnitude of $\mathbf{m}$ drops below its zero-field value, shown on Fig. \ref{fig:Mzfs} (a), 
marking the crossing point of the competition between Zeeman and spin-orbit terms.

Within the considered localized electron approach the exchange interactions do not
contribute to the ground state magnetic properties of the studied compound.
Generally, this is true for all $3d^2$ compounds since the corresponding energy gaps
are larger than $1$ eV. For example, the first singlet state in the energy spectrum
shown on Fig. \ref{fig:en} is approximately $1.846$ eV, see equation \eqref{eq:S26}.
Therefore, the only exchange related magnetic transitions occur due to absorption 
or emission.

This study provides analytical and numerical results for the expectation values 
of all possible interaction terms, energy eigenstates and the related magnetic 
moments. 
In general, the used method and the ensuing analytical expressions can 
be directly applied to all $3d^2$ single molecular magnets and 
low-dimensional systems. Some candidates, for example, are those consisting of 
Ti$^{2+}$, Cr$^{4+}$ and Mn$^{5+}$ metal centers.

	
\begin{acknowledgments}
This work was supported by the Bulgarian National Science Fund under
grant No K$\Pi$-06-H38/6 and the National Program ``Young scientists and
postdoctoral researchers'' approved by  PMC 577 / 17.08.2018.
\end{acknowledgments}

\appendix

\section{State representation}\label{app:d-state}

The single electron $3d$ states \eqref{eq:Basis_states} are
given by the product between the radial wavefunction $R_{32}(\rho)$,  
the spherical harmonics $Y^{m_l}_{2}(\theta,\phi)$ \cite{weber_essentials_2014,cohen_quantum_2020} 
and the spin states $|m_s\rangle$, where
$m_l=0,\pm1,\pm2$ and $m_s=\pm1/2$.
For example, $|d_{z^2}\rangle|m_s\rangle\to R_{32}(\rho)Y^0_2|m_s\rangle$. 
Omitting the spin term, we have 
\begin{equation*}
\begin{array}{l}
	|d_{xz}\rangle \to 
	R_{32}(\rho)\frac{1}{\sqrt{2}}
	\big(Y^{-1}_2(\theta,\phi)-Y^{1}_2(\theta,\phi)\big), \\
[0.15cm]
	|d_{yz}\rangle \to 
	R_{32}(\rho)\frac{i}{\sqrt{2}}
	\big(Y^{1}_2(\theta,\phi)+Y^{-1}_2(\theta,\phi)\big), \\
[0.15cm]
	|d_{xy}\rangle \to 
	R_{32}(\rho)\frac{i}{\sqrt{2}}
	\big(Y^{-2}_2(\theta,\phi)-Y^{2}_2(\theta,\phi)\big), \\
[0.15cm]
	|d_{x^2-y^2}\rangle \to 
	R_{32}(\rho)\frac{1}{\sqrt{2}}
	\big(Y^{2}_2(\theta,\phi)+Y^{-2}_2(\theta,\phi)\big). \\
\end{array}
\end{equation*}

\section{Exchange integrals}\label{app:exchange}

The exchange integrals $\{V_i\}^9_{i=1}$ and $\{L_1,L_2\}$ 
in \eqref{eq:Matrix_Coulomb} 
read
\begin{equation*}
\begin{array}{lll}
	V_1=\frac{42601}{483840}\frac{\gamma}{r_\mathrm{B}}Z, &  V_2=\frac{43459}{483840}\frac{\gamma}{r_\mathrm{B}}Z, &
	V_3=\frac{29731}{322560}\frac{\gamma}{r_\mathrm{B}}Z, \\
[0.15cm]
	V_4=\frac{81211}{967680}\frac{\gamma}{r_\mathrm{B}}Z, &  V_5=\frac{27833}{322560}\frac{\gamma}{r_\mathrm{B}}Z, &
	V_6=\frac{26689}{322560}\frac{\gamma}{r_\mathrm{B}}Z, \\
[0.15cm]
	V_7=\frac{5707}{967680}\frac{\gamma}{r_\mathrm{B}}Z, &  V_8=\frac{949}{322560}\frac{\gamma}{r_\mathrm{B}}Z, &
	V_9=\frac{169}{35840}\frac{\gamma}{r_\mathrm{B}}Z, \\
[0.15cm]
	L_1=\frac{3991}{483840}\frac{\gamma}{r_\mathrm{B}}Z, & 
	L_2=\frac{65}{13824}\frac{\gamma}{r_\mathrm{B}}Z. & {}
\end{array}
\end{equation*}

\section{CF integrals}\label{app:CF}

The functions $U_\alpha$ and $U_{\alpha,\beta}$, with $\alpha\in\{xz,yz,xy,x^2-y^2,z^2\}$, 
are the sum of five terms $U^{(i)}_\alpha$ and $U^{(i)}_{\alpha,\beta}$, 
respectively, where $i=1,\ldots,5$. For example, $U_{yz}=\sum^5_{i=1}U^{(i)}_{yz}$.

For all $i$, the functions that enter into the diagonal elements read
\begin{align*}
	U^{(i)}_{xz}=
	&
	\tfrac{\gamma Z_i}{32\varrho^5_i Z^4}
	\Big(
-21870r_\mathrm{B}^4 + 144 \varrho_i^2r_\mathrm{B}^2 Z^2 + 32 \varrho_i^4 Z^4 
+ 54r_\mathrm{B}^2 
\\ &
\times\!(675r_\mathrm{B}^2 + 8 \varrho_i^2 Z^2)\cos[2 \varphi_i] 
- 42525r_\mathrm{B}^4\!\cos[2 (\varphi_i\! -\!2\vartheta_i)] 
\\ &
+ 24300r_\mathrm{B}^4\! \cos[2 (\varphi_i\! - \vartheta_i)] - 
216 \varrho_i^2r_\mathrm{B}^2 Z^2\! \cos[2 (\varphi_i\! - \vartheta_i)] 
\\ &
- 48600r_\mathrm{B}^4 \cos[2 \vartheta_i] + 
432 \varrho_i^2r_\mathrm{B}^2 Z^2 \cos[2 \vartheta_i] - 85050r_\mathrm{B}^4 
\\ &
\times\cos[4 \vartheta_i] + 24300r_\mathrm{B}^4 \cos[2 (\varphi_i + \vartheta_i)] 
- 216 \varrho_i^2r_\mathrm{B}^2 Z^2 
\\ &
\times\cos[2 (\varphi_i + \vartheta_i)]- 42525r_\mathrm{B}^4 \cos[2 (\varphi_i + 2 \vartheta_i)]
	\Big),
\end{align*}
\begin{align*}
	U^{(i)}_{yz}=
	&
	\tfrac{\gamma Z_i}{32\varrho^5_i Z^4}
	\Big(
-21870r_\mathrm{B}^4 + 144 \varrho_i^2r_\mathrm{B}^2 Z^2 + 32 \varrho_i^4 Z^4 - 
54r_\mathrm{B}^2 
\\ &
\times\!(675r_\mathrm{B}^2 + 8 \varrho_i^2 Z^2)\cos[2 \varphi_i] + 
42525r_\mathrm{B}^4\!\cos[2(\varphi_i\! -\!2\vartheta_i)] 
\\ &
- 24300r_\mathrm{B}^4\!\cos[2 (\varphi_i\! - \vartheta_i)] + 
216 \varrho_i^2r_\mathrm{B}^2 Z^2\!\cos[2(\varphi_i\! - \vartheta_i)] 
\\ &
- 48600r_\mathrm{B}^4\cos[2 \vartheta_i] + 
432 \varrho_i^2r_\mathrm{B}^2 Z^2\cos[2 \vartheta_i] - 85050r_\mathrm{B}^4
\\ &
\times\cos[4 \vartheta_i] - 
24300r_\mathrm{B}^4\cos[2 (\varphi_i + \vartheta_i)] + 216 \varrho_i^2r_\mathrm{B}^2 Z^2
\\ &
\times\cos[2 (\varphi_i + \vartheta_i)] + 
42525r_\mathrm{B}^4\cos[2 (\varphi_i + 2 \vartheta_i)]
	\Big),
\end{align*}
\begin{align*}
	U^{(i)}_{xy}=
	&
	\tfrac{\gamma Z_i}{128\varrho^5_i Z^4}
	\Big(
21870r_\mathrm{B}^4 - 1152\varrho_i^2r_\mathrm{B}^2 Z^2\! + 128 \varrho_i^4 Z^4\! - 
255150
\\ &
\times r_\mathrm{B}^4\cos[4\varphi_i] + 170100r_\mathrm{B}^4\cos[4\varphi_i - 2 \vartheta_i] - 
42525r_\mathrm{B}^4
\\ &
\times\cos[4 (\varphi_i - \vartheta_i)] + 48600r_\mathrm{B}^4\cos[2 \vartheta_i] 
- 3456 \varrho_i^2r_\mathrm{B}^2 Z^2
\\ &
\times\cos[2 \vartheta_i] + 85050r_\mathrm{B}^4\cos[4 \vartheta_i] - 
42525r_\mathrm{B}^4
\\ &
\times\cos[4 (\varphi_i + \vartheta_i)] + 170100r_\mathrm{B}^4\cos[2 (2 \varphi_i + \vartheta_i)]
	\Big),
\end{align*}
\begin{align*}
	U^{(i)}_{x^2-y^2} \!\! =
	&
	\tfrac{\gamma Z_i}{128\varrho^5_i Z^4}
	\Big( 
21870r_\mathrm{B}^4 -\! 1152 \varrho_i^2r_\mathrm{B}^2 Z^2\! + 128 \varrho_i^4 Z^4 \! + \!255150
\\ &
\times r_\mathrm{B}^4\cos[4 \varphi_i] - 170100r_\mathrm{B}^4\cos[4 \varphi_i\! - 2 \vartheta_i] +\! 
42525r_\mathrm{B}^4
\\ &
\times\cos[4 (\varphi_i\! - \vartheta_i)] + 48600r_\mathrm{B}^4\!\cos[2 \vartheta_i] -\! 
3456 \varrho_i^2r_\mathrm{B}^2 Z^2
\\ &
\times\cos[2 \vartheta_i] + 85050r_\mathrm{B}^4\cos[4 \vartheta_i] + 
42525r_\mathrm{B}^4
\\ &
\times\cos[4 (\varphi_i + \vartheta_i)] - 170100r_\mathrm{B}^4\cos[2 (2 \varphi_i + \vartheta_i)]
	\Big)
\end{align*}
and
\begin{align*}
	U^{(i)}_{z^2}=
	&
	\tfrac{\gamma Z_i}{32\varrho^5_i Z^4}
	\Big( 
	32805r_\mathrm{B}^4 + 288 \varrho_i^2r_\mathrm{B}^2 Z^2 + 32 \varrho_i^4 Z^4 + 
	108r_\mathrm{B}^2 
	\\ &
	\times(675r_\mathrm{B}^2 + 8 \varrho_i^2 Z^2)\cos[2 \vartheta_i] + 
	127575r_\mathrm{B}^4\cos[4 \vartheta_i]
	\Big).
\end{align*}
Furthermore, $\forall \ i$, for the functions \eqref{eq:Matrix_CFoff}
we have 
\begin{align*}
	U^{(i)}_{xz,yz} =
	&
	\tfrac{27\gamma r^2_\mathrm{B} Z_i}{4\varrho^5_i Z^4}
	\Big( 
		1125r_\mathrm{B}^2 + 4 \varrho_i^2 Z^2 + 1575r_\mathrm{B}^2\cos[2 \vartheta_i]
\\  &
		\times\sin[2 \varphi_i]\sin^2[\vartheta_i]
	\Big),
\end{align*}
\begin{align*}
	U^{(i)}_{xz,xy} =
	&
	\tfrac{27\gamma r^2_\mathrm{B} Z_i}{16\varrho^5_i Z^4}
	\Big( 
6300r_\mathrm{B}^2\cos[\vartheta_i]\sin[3 \varphi_i]\sin^3[\vartheta_i] 
\\  &
- \big(225r_\mathrm{B}^2 - 16 \varrho_i^2 Z^2 + 1575r_\mathrm{B}^2\cos[2 \vartheta_i]\big)
\\ &
\times\sin[\varphi_i]\sin[2 \vartheta_i]
	\Big),
\end{align*}
\begin{align*}
	U^{(i)}_{xz,x^2-y^2} =
	&
	\tfrac{27\gamma r^2_\mathrm{B} Z_i}{32\varrho^5_i Z^4}
	\Big( 
\big(8 \varrho_i^2 Z^2 + 1575r_\mathrm{B}^2\cos[2 \varphi_i]-900r_\mathrm{B}^2\big)
\\ &
\times4\cos[\varphi_i]\sin[2 \vartheta_i] - 
1575r_\mathrm{B}^2 
\\ &
\times\big(\cos[\varphi_i]+\cos[3 \varphi_i]\big)\sin[4 \vartheta_i]
	\Big),
\end{align*}
\begin{align*}
	U^{(i)}_{xz,z^2} =
	&
	\tfrac{9\sqrt{3}\gamma r^2_\mathrm{B} Z_i}{8\varrho^5_i Z^4}
	\big(675r_\mathrm{B}^2 + 8 \varrho_i^2 Z^2 + 4725r_\mathrm{B}^2\cos[2 \vartheta_i]
	\big)
\\ &	
	\times\cos[\varphi_i]\sin[2 \vartheta_i],
\end{align*}
\begin{align*}
	U^{(i)}_{yz,xy} =
	&
	\tfrac{27\gamma r^2_\mathrm{B} Z_i}{8\varrho^5_i Z^4}
	\Big(
\big(675r_\mathrm{B}^2 + 8\varrho_i^2 Z^2 - 1575r_\mathrm{B}^2\cos[2 \varphi_i]\big) 
\\ &
\times\sin[2 \vartheta_i]-1575r_\mathrm{B}^2\sin^2[\varphi_i]\sin[4 \vartheta_i]
	\Big)\cos[\varphi_i],
\end{align*}
\begin{align*}
	U^{(i)}_{yz,x^2-y^2} =
	&
	\tfrac{27\gamma r^2_\mathrm{B} Z_i}{2\varrho^5_i Z^4}
	\cos[\vartheta_i]\sin[\varphi_i]\sin[\vartheta_i]
\\ &
	\times\big(
	450r_\mathrm{B}^2-4\varrho_i^2 Z^2 +\! 1575r_\mathrm{B}^2\cos[2 \varphi_i]\sin^2[\vartheta_i]
	\big),
\end{align*}
\begin{align*}
	U^{(i)}_{yz,z^2} =
	&
	\tfrac{9\sqrt{3}\gamma r^2_\mathrm{B} Z_i}{8\varrho^5_i Z^4}
	\big(675r_\mathrm{B}^2 + 8 \varrho_i^2 Z^2 + 4725r_\mathrm{B}^2
	\cos[2 \vartheta_i]
	\big)
	\\ &	
	\times\sin[\varphi_i]\sin[2 \vartheta_i],
\end{align*}
\begin{align*}
	U^{(i)}_{xy,x^2-y^2} =
	&
	\tfrac{42525\gamma r^4_\mathrm{B} Z_i}{8\varrho^5_i Z^4}
	\sin[4\varphi_i]\sin^4[\vartheta_i],
\end{align*}
\begin{align*}
	U^{(i)}_{xy,z^2} =
	&
	\tfrac{9\sqrt{3}\gamma r^2_\mathrm{B} Z_i}{8\varrho^5_i Z^4}
	\big(3375r_\mathrm{B}^2-16 \varrho_i^2 Z^2 + 4725r_\mathrm{B}^2
	\cos[2 \vartheta_i]\big)
\\ &	
	\times\sin[2\varphi_i]\sin^2[\vartheta_i]
\end{align*}
and
\begin{align*}
	U^{(i)}_{x^2-y^2,z^2} =
	&
	\tfrac{9\sqrt{3}\gamma r^2_\mathrm{B} Z_i}{8\varrho^5_i Z^4}
	\big(3375r_\mathrm{B}^2-16 \varrho_i^2 Z^2 + 4725r_\mathrm{B}^2
	\cos[2 \vartheta_i]\big)
	\\ &	
	\times\cos[2\varphi_i]\sin^2[\vartheta_i].
\end{align*}

\section{Energy spectrum}\label{app:enspectrum}

The zero-field values, in eV, of all energy levels depicted on Fig. \ref{fig:en} (a) are 
provided hereafter. Starting from the highest one on the left to the ground 
state energy level on the right, we have the array
\begin{align*}
	\big\{
	& 5.81233, 3.90346, 3.75779, 3.32676, 3.16796, 2.92855, 
\\	& 2.92845, 2.92835, 2.8337, 2.7857, 2.78561, 2.7854, 2.77113, 
\\	& 2.43523, 2.40827, 2.32306, 2.13117, 1.98332, 1.91452,  
\\	& 1.84694, 1.82376, 1.82358, 1.82344, 1.6839, 1.68375, 
\\	& 1.68362, 1.6435, 1.19804, 1.19803, 1.19803, 0.778733,   
\\  & 0.778659, 0.778587,0.287733, 0.287647, 0.287533,    
\\  & 0.212738, 0.212734, 0.212714, 0.204263, 0.204227,  
\\ 	& 0.204211, 0.000364953, 0.000170839, 0\big\}.
\end{align*}

The second excited state in the zero-field energy spectrum depicted on Fig. 
\ref{fig:en} (b) reads 
\begin{align}\label{eq:Ap_ZExstate_2}
	|\Psi_3\rangle=&
	(-0.16 + 0.14i) |\psi_{1,1,0}\rangle+(0.13 + 0.15i)|\psi_{1,1,\bar{1}}\rangle 
	\nonumber \\ & 
	+(0.28 - 0.24i)|\psi_{2,1,0}\rangle - (0.22 + 0.26i) |\psi_{2,1,\bar{1}}\rangle 
	\nonumber \\ & 
	-(0.15 - 0.12i)|\psi_{3,1,0}\rangle + (0.11 + 0.14i) |\psi_{3,1,\bar{1}}\rangle 
	\nonumber \\ & 
	-(0.27 - 0.23i)|\psi_{5,1,0}\rangle + (0.22 + 0.25i) |\psi_{5,1,\bar{1}}\rangle 
	\nonumber \\ & 
	+(0.33 - 0.28i)|\psi_{6,1,0}\rangle - (0.25 + 0.31i)|\psi_{6,1,\bar{1}}\rangle 
	\nonumber \\ & 
	- (0.074 - 0.063i)|\psi_{8,1,0}\rangle + (0.058 +0.07i)|\psi_{8,1,\bar{1}}\rangle
	\nonumber \\ &
	+ \sum_{n\ge2} O\big(10^{-n}\big)|\psi_{\ldots}\rangle.
\end{align}
In the case shown on Fig. \ref{fig:en} (c), the state
\eqref{eq:Ap_ZExstate_2} transforms into
\begin{align}\label{eq:Ap_Exstate_2}
	|\Psi^\mathrm{f}_3\rangle=&
	(-0.046 + 0.043i) |\psi_{1,1,0}\rangle+(0.19 + 0.21i)|\psi_{1,1,\bar{1}}\rangle 
	\nonumber \\ & 
	+(0.080 - 0.076i)|\psi_{2,1,0}\rangle - (0.32 + 0.36i) |\psi_{2,1,\bar{1}}\rangle 
	\nonumber \\ & 
	-(0.042 - 0.038i)|\psi_{3,1,0}\rangle + (0.16 + 0.20i) |\psi_{3,1,\bar{1}}\rangle 
	\nonumber \\ & 
	-(0.077 - 0.072i)|\psi_{5,1,0}\rangle + (0.32 + 0.35i) |\psi_{5,1,\bar{1}}\rangle 
	\nonumber \\ & 
	+(0.096 - 0.087i)|\psi_{6,1,0}\rangle - (0.38 + 0.43i)|\psi_{6,1,\bar{1}}\rangle 
	\nonumber \\ & 
	- (0.022 - 0.02i)|\psi_{8,1,0}\rangle + (0.086 + 0.097i)|\psi_{8,1,\bar{1}}\rangle
	\nonumber \\ &
	+ \sum_{n\ge2} O\big(10^{-n}\big)|\psi_{\ldots}\rangle.
\end{align}

\section{Photoluminescence}\label{app:photo}

For the explicit representation of the $15$-th energy state, 
related to the emission 
with $1240$ nm wavelength obtained in the case $\Delta\vartheta_i\to0.585^\circ$ 
and $\Delta\varphi_i=0^\circ$ for all $i=1,\ldots,5$, we have
\begin{align*}
	|\Psi^{\Delta}_{15}\rangle\! =&
	(0.34 + 0.56i)|\psi_{1,1,0}\rangle - (0.55 - 0.35i)|\psi_{1,1,\bar{1}}\rangle 
	\\ &
	+ (0.11 + 0.18i)|\psi_{2,1,0}\rangle - (0.17 - 0.11i)|\psi_{2,1,\bar{1}}\rangle 
	\\ &
	- (0.022\! + 0.036i)|\psi_{5,1,0}\rangle + (0.036\! -0.022i)|\psi_{5,1,\bar{1}}\rangle 
	\\ &
	+ (0.044\! + 0.071i)|\psi_{6,1,0}\rangle - (0.069\! -0.044i)|\psi_{6,1,\bar{1}}\rangle 
	\\ & 
	- (0.073\! + 0.12i)|\psi_{8,1,0}\rangle + (0.117\! -0.074i)|\psi_{8,1,\bar{1}}\rangle 
	\\ &
	+ \sum_{n\ge2} O\big(10^{-n}\big)|\psi_{\ldots}\rangle.
\end{align*}
To compare the last superposition with that in the case of no change in 
the ligands angles, see \eqref{eq:S15}.



%

\end{document}